\documentclass[twocolumn]{aastex63}

\usepackage{gensymb}
\usepackage{bm}
\usepackage{tabularx}

\revised{\today}
\shorttitle{Earth Through Time}
\shortauthors{Goodis Gordon et al. 2025}
\graphicspath{{./}{figures/}}

\begin{document}

\title{Polarized Signatures of the Earth Through Time:\\An Outlook for the Habitable Worlds Observatory}

\correspondingauthor{Kenneth E. Goodis Gordon}
\email{Kenneth.Gordon@ucf.edu}

\author[0000-0002-4258-6703]{Kenneth E. Goodis Gordon}
\affiliation{Planetary Sciences Group, Department of Physics, University of Central Florida, 4111 Libra Drive, Orlando, FL 32816, USA}

\author[0000-0001-7356-6652]{Theodora Karalidi}
\affiliation{Planetary Sciences Group, Department of Physics, University of Central Florida, 4111 Libra Drive, Orlando, FL 32816, USA}

\author[0000-0002-4420-0560]{Kimberly M. Bott}
\affiliation{SETI Institute, Mountain View, CA 94043, USA}
\affiliation{NASA Nexus for Exoplanet System Science, Virtual Planetary Laboratory Team, Box 351580, University of Washington, Seattle, WA 98195, USA}

\author[0000-0002-0413-3308]{Nicholas F. Wogan}
\affiliation{Space Science Division, NASA Ames Research Center, Moffett Field, CA 94035, USA}
\affiliation{NASA Nexus for Exoplanet System Science, Virtual Planetary Laboratory Team, Box 351580, University of Washington, Seattle, WA 98195, USA}

\author[0000-0001-6285-267X]{Giada N. Arney}
\affiliation{NASA Goddard Space Flight Center, 8800 Greenbelt Road, Greenbelt, MD 20771, USA}
\affiliation{GSFC Sellers Exoplanet Environments Collaboration, NASA Goddard Space Flight Center, Greenbelt, MD 20771, USA}
\affiliation{NASA Nexus for Exoplanet System Science, Virtual Planetary Laboratory Team, Box 351580, University of Washington, Seattle, WA 98195, USA}

\author[0000-0003-1225-6727]{Mary N. Parenteau}
\affiliation{Exobiology Branch, NASA Ames Research Center, Moffett Field, CA 94035, USA}

\author[0000-0003-3759-9080]{Tiffany Kataria}
\affiliation{Astrophysics Section, NASA Jet Propulsion Laboratory, California Institute of Technology, Pasadena, CA 91109, USA}

\author[0000-0002-1386-1710]{Victoria S. Meadows}
\affiliation{Department of Earth and Space Sciences/Astrobiology Program, University of Washington, Seattle, WA 98195, USA}
\affiliation{NASA Nexus for Exoplanet System Science, Virtual Planetary Laboratory Team, Box 351580, University of Washington, Seattle, WA 98195, USA}



\begin{abstract}

The search for life beyond the Solar System remains a primary goal of current and near-future missions, including NASA's upcoming Habitable Worlds Observatory (HWO). However, research into determining the habitability of terrestrial exoplanets has been primarily focused on comparisons to modern-day Earth. Additionally, current characterization strategies focus on the unpolarized flux from these worlds, taking into account only a fraction of the informational content of the reflected light. Better understanding the changes in the reflected light spectrum of the Earth throughout its evolution, as well as analyzing its polarization, will be crucial for mapping its habitability and providing comparison templates to potentially habitable exoplanets. Here we present spectropolarimetric models of the reflected light from the Earth at six epochs across all four geologic eons. We find that the changing surface albedos and atmospheric gas concentrations across the different epochs allow the habitable and non-habitable scenarios to be distinguished, and diagnostic features of clouds and hazes are more noticeable in the polarized signals. We also discuss how using Mie scattering for naturally non-spherical particles, which is a common simplification for exoplanet modeling, affects the resulting planetary signals.
Finally, our results suggest that pushing the HWO planet-to-star flux contrast limit down to 1 $\times$ 10$^{-13}$ could allow for the characterization in both unpolarized and polarized light of an Earth-like planet at any stage in its history.

\end{abstract}



\section{Introduction} \label{sec:intro}

Recent technological advancements and improved observational capabilities have allowed for the detection of thousands of exoplanets, including approximately 200 terrestrial planets \citep[e.g.,][]{akeson2013, NASAExo}. Missions including Kepler and TESS have enabled observations of dozens of rocky exoplanets in the habitable zones (HZs) of their stars \citep[e.g.,][]{batalha2013, torres2015, kane2016, hill2023}. The recently launched James Webb Space Telescope (JWST) and the developing instruments for the Extremely Large Telescopes (ELTs) \citep[e.g.,][]{wright2016, thatte2021, males2022} will continue the search for these rocky planets in the near future.

The next major step lies in their characterization, particularly, identifying biosignatures and determining the habitability of these worlds. This is a complex problem and a range of different planetary, orbital, and stellar parameters need to be taken into account. To date, Earth is the only planet known to harbor life. Therefore, Earth is the benchmark from which we infer the biosignatures of a habitable planet. However, existing studies of the habitability of exoplanets have so far focused mainly on comparisons to modern-day Earth \citep[e.g.,][]{sagan1993, woolf2002, robinson2011, kopparapu2013, feng2018},
even though Earth's atmosphere and surface have undergone significant evolution since its formation.
Both empirical biogeochemical analyses as well as theoretical studies provide examples of past Earth that seem alien in nature yet were still habitable \citep[for some overviews, see, e.g.,][]{schwieterman2018, robinson2018}.

Some past research has simulated the changes in Earth's reflection, emission, and transmission spectra across geologic time. For example, 
\citet[][]{kaltenegger2007} used atmospheric concentration profiles to model the reflection and emission spectra of Earth for six long-lived periods of its history, ranging from 3.9 Ga to the present. Focusing only on the most spectrally active species, they analyzed cloud, surface, and atmospheric contributions on the spectra throughout the different periods and determined the resolutions required to adequately detect the main features for each epoch.
Additional work by members of the same group extended this study to investigate the stellar contributions on the spectra if the Earth were orbiting host stars of different spectral types \citep[][]{rugheimer2018}. However, these studies did not include any models of the Earth in its first few hundred million years and ignored relatively short-term events such as glaciation events or hothouse-Earth events.

The models of \citet[][]{kaltenegger2007} and \citet[][]{rugheimer2018} also did not include any haze in their simulated atmospheres. In reality, the anoxic atmosphere of the early Earth may have supported the formation of an organic haze \citep[e.g.,][]{pavlov2001haze, wolf2010, zerkle2012, claire2014}, which could have either heated or cooled the globe \citep[e.g.,][]{mak2023}. \citet[][]{arney2016, arney2017, arney2018} modeled the so-called ``Pale Orange Dots" and investigated the impact of hydrocarbon haze on the early Earth’s habitability and surface temperature. \citet[][]{arney2016} also studied early Earth atmospheres with varying levels of O$_2$ to determine how different oxygen concentrations in those early atmospheres could create ozone layers similar to hazes that could potentially shield the surface from harmful ultraviolet radiation. However, their models only included ocean or icy surfaces and were only generated for planets at quadrature (i.e., the planets are half illuminated with respect to the observer).

Recent studies by \citet[][]{wogan2020} and \citet[][]{zahnle2020} modeled short-term events for the Earth in its earliest stages. \citet[][]{zahnle2020} explored how different sized impactors could have transiently created H$_2$-rich atmospheres early in Earth's history.
\citet[][]{wogan2020} provided the first estimates of chemical disequilibrium during Earth's earliest eon, the Hadean, and investigated when disequilibrium might indicate life versus when disequilibrium serves as an anti-biosignature. Although these studies provide important context for understanding the earliest atmospheres of Earth, they do not provide any simulated spectra of the Earth for these time periods.

While these previous modeling efforts may aid the interpretation of future observations of Earth-like planets, their simulations only made use of the unpolarized flux from the planets, thereby losing some of the informational content available from the light. Polarimetry, on the other hand, measures light as a vector rather than just a scalar flux intensity and allows for the use of 100\% of the informational content of the light. Polarimetry can improve the accuracy of flux simulations even when polarization is not of interest, and studies have shown that neglecting polarization in these simulations can result in errors of up to a few tens of percent \citep[e.g.,][]{mishchenko1994, stam2005, emde2018}.

Spectropolarimetry can provide detailed information about the physical mechanism scattering the light, thereby allowing for accurate characterizations of the properties of a planetary atmosphere and surface, including its chemical composition, thermal structure, cloud particle size, cloud top pressure, and surface albedo \citep[e.g.,][]{hansenhovenier1974, hansentravis1974, trees2022, gordon2023}. The vector nature of polarimetry also makes it extremely sensitive to the location of specific features on the observed disk of the object \citep[see e.g.,][]{karalidi2013flux, stolker2017}. Polarimetry therefore has the ability to break degeneracies of flux-only observations \citep[e.g.,][]{hansenhovenier1974, fauchez2017, rossistam2017}. Studies on the spectropolarization of the earthshine revealed diagnostic biosignatures of the Earth, including the Vegetation Red Edge (VRE) and spectral features of key atmospheric gases, in addition to showing the sensitivity of polarization to features such as water clouds, varying surfaces, and ocean glint \citep[e.g.,][]{sterzik2012, sterzik2019, sterzik2020, bazzon2013, takahashi2013, takahashi2021, miles2014, gordon2023}.

To date, no analyses of the polarization of reflected light from early-Earth analogs exist. Here we utilized an advanced polarization-enabled radiative transfer code to model the unpolarized and polarized visible to near-infrared (VNIR) reflected flux of the Earth, as functions of both wavelength $\lambda$ and planetary phase angle $\alpha$, across all four geologic eons, including the first models of the spectra of the Hadean Earth. Our models cover both short-term and long-term periods in Earth's history and include atmospheric and surface profiles from numerous studies in the literature. All of our models assume a planet with the same surface gravity as Modern Earth (9.81 m s$^{-2}$) orbiting 1 AU from an evolving Sun. Our models are publicly available online at \dataset[doi:10.5281/zenodo.13882511]{\doi{10.5281/zenodo.13882511}} for the community to help inform instrument design decisions and to assist in characterizing Earth-like planets.

The outline of this paper is as follows. In Section~\ref{sec: Defs} we give general descriptions of reflected light polarization and present the radiative transfer code used in our study. Section~\ref{sec:Models} provides descriptions of the atmospheric, surface, cloud, and haze properties we use in our models and presents justifications for the inputs for our chosen Earth epochs. In Section~\ref{sec: Results}, we present our resulting flux and polarization spectra, highlighting the effects of the clouds and hazes on the models. In Section~\ref{sec: Realistic} we explore how the addition of more physically consistent parameters for the clouds, hazes, and surfaces affect the resulting polarization. In Section~\ref{sec: Scaling} we provide first-order observing constraints for upcoming polarimeters aimed at characterizing terrestrial exoplanets. Finally, in Section~\ref{sec: DiscussConclude} we discuss and summarize our results and present a future outlook.

\section{Calculating the Polarization of Reflected Light} \label{sec: Defs}

\subsection{Defining Flux and Polarization} \label{sec: defs}

Starlight that has been reflected by a planet can be fully described by its flux vector $\pi\textbf{F}$ \citep[see, e.g.,][]{hansentravis1974, hovenier2004}, as

\begin{eqnarray}
\pi\textbf{F}  =  \pi\left[\begin{array}{c} F \\
Q \\
U \\
V\end{array}\right]
\label{eq:fluxvec}
\end{eqnarray}
where parameter $\pi$F is the total reflected flux, parameters $\pi$Q and $\pi$U are the linearly polarized fluxes, and parameter $\pi$V is the circularly polarized flux. All four parameters are wavelength-dependent and their units are in $W$ $m^{-2} m^{-1}$ when defined per wavelength. 
The fluxes are defined in reference to the planetary scattering plane (i.e., the plane through the centers of the host star, planet, and observer; for more details see, e.g., \citet[][]{stam2008, gordon2023}).

The total degree of polarization, P$_{tot}$, of the light that is reflected by a planet is defined as the ratio of the polarized fluxes to the total flux, as:

\begin{eqnarray}
P_{tot} = \frac{\sqrt{Q^{2} + U^{2} + V^{2}}}{F}.
\label{eq:Ptot}
\end{eqnarray}

Studies have shown that parameter $\pi$V of reflected sunlight from an Earth-like planet will be negligible \citep[e.g.,][]{hansentravis1974, rossi2018}, and ignoring it does not lead to any significant errors in the calculated total and polarized fluxes \citep{stam2005}. Therefore, we ignored it in our simulations here. 
Additionally, for a planet that is mirror-symmetric with respect to the planetary scattering plane (i.e., horizontally homogeneous), parameter $\pi$U will be effectively zero \citep[e.g.,][]{hovenier1970}. In this case, we can define the \textit{signed} degree of linear polarization, which also includes the direction of the polarization, as \citep[see also, e.g.,][]{gordon2023}:

\begin{eqnarray}
P_{s} & = & \frac{-Q}{F}.
\label{eq:degofpol}
\end{eqnarray}
If P$_{s} > 0$, the light is polarized perpendicular to the planetary scattering plane, whereas if P$_{s} < 0$, the light is polarized parallel to the plane.

\subsection{The Radiative Transfer Code} \label{sec: codes}

To generate the synthetic unpolarized and polarized signatures of our model planets, we used the Doubling Adding Program (DAP) polarization-enabled radiative transfer code. 
DAP fully incorporates single and multiple scattering by atmospheric gases as well as aerosol and cloud particles and can model atmospheres of any composition with as many layers as needed to describe the full scattering properties of the atmosphere.

Our version of DAP uses an efficient adding-doubling algorithm \citep[][]{hansentravis1974, deHaan1987} coupled with a fast, numerical disk integration routine \citep[][]{stam2006}. 
Our code uses the HITRAN 2020 molecular line lists \citep[][]{gordon2022} and the k-coefficient method to calculate the absorption properties of the atmospheric gases. For a discussion of the effects of using k-coefficients versus line-by-line calculations on polarized model spectra, we refer the reader to \citet[][]{gordon2023}. This study benchmarked our version of DAP against another polarization-enabled radiative transfer code VSTAR \citep[][]{bailey2018}, which makes use of the VLIDORT code of \citet[][]{spurr2006} and uses line-by-line calculations, and showed that results from the two codes were in general agreement with each other.
Different versions of DAP over the years have been used to calculate the flux and polarization signals of both terrestrial and gaseous planets \citep[e.g.,][]{stam2006, stam2008, dekok2011, karalidi2011, karalidi2012rainbow, karalidistam2012, karalidi2013flux, fauchez2017, rossistam2017, rossi2018, treesstam2019, groot2020, trees2022, gordon2023, mahapatra2023, chubb2024}.

DAP defines the flux vector of stellar light that has been reflected by a spherical planet with radius $r$ at a distance $d$ from the observer (where $d \gg r$) as:

\begin{eqnarray}
\pi\textbf{F}(\lambda, \alpha)  =  \frac{1}{4}\frac{r^2}{d^2}\textbf{S}(\lambda, \alpha)\pi\textbf{F}_0(\lambda)
\label{eq:fluxplanet}
\end{eqnarray}
where $\lambda$ is the wavelength of the light and $\alpha$ is the planetary phase angle. $\pi\textbf{F}_0$ is the flux vector of the incident starlight and $\textbf{S}$ is the $4 \times 4$ planetary scattering matrix with elements $a_{ij}$, which is calculated internally in DAP \citep[for more information see][]{stam2006}.
For our calculations, we normalized Equation \ref{eq:fluxplanet} assuming $r = 1$, $d = 1$, and unpolarized incident starlight \citep[e.g.,][]{kemp1987, cotton2017} so that $\pi{F_{0}} = 1$. The total reflected flux then becomes:

\begin{eqnarray}
\pi{F_{n}}(\lambda, \alpha)  =  \frac{1}{4}a_{11}(\lambda, \alpha)
\label{eq:normalizedflux}
\end{eqnarray}
where $a_{11}$ is the (1,1)-element of the $\textbf{S}$ matrix and the subscript $n$ on the flux indicates that it is now normalized. $\pi{F_{n}}$ corresponds to the planet's geometric albedo $A_G$ when $\alpha$ = 0$\degree$ \citep[see, e.g.,][]{stam2008}. These normalized fluxes can be straightforwardly scaled for any planetary system using Equation \ref{eq:fluxplanet} and inserting the correct values for $r$, $d$, and $\pi{F_{0}}$. In Section \ref{sec: Scaling} we provide preliminary constraints for observed fluxes of our Earth Through Time models.

\section{Model Descriptions} \label{sec:Models}

Earth has gone through multiple stages of habitability and non-habitability throughout its evolution, over both long (i.e., multiple geological eras) and short (i.e., individual geological periods) timescales. Here we chose to model six different epochs, including three short- and three long-term epochs, of our planet’s history which exhibit a range of atmospheric compositions, temperature-pressure (T-P) profiles, and surface characteristics. To make our models as realistic as possible and to capture the horizontal inhomogeneity of the visible Earth disk, we divided each model planet into pixels with locally plane-parallel, horizontally homogeneous, and vertically heterogeneous atmosphere and surface properties. 
We then ran our radiative transfer code (see Section~\ref{sec: codes}) for each unique pixel combination to produce the full spectropolarimetric signal of the pixel for wavelengths ($\lambda$) between 0.3 and 1.8~$\mu$m and all phase angles ($\alpha$) from $0\degree - 180\degree$. Our models have a constant spectral resolution ($\Delta\lambda$) of 10 nm, corresponding to spectral resolving powers of R $(= \lambda / \Delta\lambda) \sim 50$ in the visible (R$_{VIS}$) and R $\sim 150$ in the NIR (R$_{NIR}$), and cover the full phase space in steps of $\alpha$ = 2\degree. Spectra for the horizontally inhomogeneous planets were then simulated using the weighted sum approximation \citep[see e.g.,][]{stam2008, karalidistam2012}. The flux vector of a planet covered by $M$ different types of pixels is calculated as:  $\pi\textbf{F}(\alpha) = \sum_{m = 1}^{M} w_{m} \pi\textbf{F}_{m}(\alpha)$, where $\pi\textbf{F}_m$ is the flux vector from a single $m$ pixel and $w_{m}$ is the fraction of type $m$ pixels on the inhomogeneous planet, so that $\sum_{m = 1}^{M} w_{m} = 1$.

\subsection{Model Atmospheres} \label{sec: Atmos}

All model pixels have vertically heterogeneous atmospheres containing gas molecules and (if desired) cloud or haze particles. Each atmosphere is bounded below by a flat, homogeneous surface (see Section~\ref{sec: Surfs}). To simulate the Hadean Earth atmospheres, we used the 1-D photochemical-climate code \textit{Photochem} \citep[e.g.,][]{wogan2023}, while for the Archean, Proterozoic, and Modern eons we used the \textit{Atmos} code \citep[see, e.g.,][and references therein]{arney2016, arney2017, arney2018}. Both \textit{Photochem} and \textit{Atmos} solve the 1-D continuity equation \citep[e.g., see Appendix B in][]{catling2017} describing vertical gas and particle transport, chemical reactions, condensation, and rainout in droplets of water. Our \textit{Photochem} calculations used the \citet[][]{wogan2023} chemical network, while our \textit{Atmos} models used the network described in \citet[][]{arney2016}.

Our outputs from \textit{Photochem} and \textit{Atmos} originally consisted of 100 (for all Hadean Earth epoch models) or 200 (for the Archean, Proterozoic, and Modern Earth epoch models) atmospheric layers. 
To maximize the computational storage efficiency of this study, we performed an analysis to calculate the ideal number of atmospheric layers to use for our models. We found that 45 layers were sufficient to represent our model atmospheres without any significant change (i.e., negligible change in the continua and $<$1\% absolute difference in the deepest absorption bands) in the resulting polarization of the reflected light.

\subsection{Model Clouds and Hazes} \label{sec: CloudsHazes}

We modeled both clear (i.e., cloud- and haze-free) as well as cloudy or hazy atmospheres. In the latter, one or more atmospheric layers contain cloud or haze particles in addition to the gas molecules.
Following \citet[][]{stam2008} and \citet[][]{gordon2023}, we then used the weighted averaging method to model the signals of horizontally heterogeneous exoplanets with patchy clouds or hazes.

For our cloudy atmosphere models, we analyze two different modern Earth cloud cases: 
liquid water droplets that are representative of Stratocumulus clouds, and water ice particles that are representative of Cirrus clouds. Both clouds are placed in the appropriate atmospheric layer corresponding to an altitude of $\sim$1 km and $\sim$10 km, respectively. 
Optical thicknesses, $\tau$, of both liquid and ice water clouds on Earth show large variations across time and location. 
For simplicity, the optical thicknesses of our Stratocumulus clouds are set to $\tau$ = 10 and those of our Cirrus clouds are set to $\tau$ = 0.5 for all models, based on cloud cover properties derived from observations by the Moderate Resolution Imaging Spectroradiometer (MODIS) instrument on-board NASA’s Terra and Aqua satellites \citep[e.g.,][]{king2004, king2013}.

For our hazy atmosphere models, we modeled hydrocarbon haze particles. Unlike with our liquid and ice water clouds, whose properties are constant for all models, the particle radii and number density per atmospheric layer of our hazes are based on the \textit{Photochem} and \textit{Atmos} models (for more information on these hazes, including their shapes and properties, we refer the reader to \citet[][]{arney2016, wogan2023}). For simplicity, we binned the haze particle radii generated by \textit{Photochem} and \textit{Atmos} into six particle radii ``modes” ranging from 0.01 $\mu$m to 0.75 $\mu$m. After binning the particle radii, we computed the total $\tau$ of each mode at a reference wavelength of 0.55 $\mu$m in each atmospheric layer.
For haze particles of a given radius, R, $\tau$ depends on the thickness of the atmospheric layer, z, the wavelength-dependent extinction efficiency, $Q_{ext}$, and the number density of the haze particles in the layer, $n_{R}$:

\begin{eqnarray}
\tau(R) = z \times \pi R^{2} \times Q_{ext}(\lambda,R) \times n_{R}.
\label{eq:taus}
\end{eqnarray}
\citep[see, e.g.,][]{arney2016}

For our simulations we modeled all cloud and haze particles as spheres. We modeled the optical properties of these aerosols, including their Q$_{ext}$, using Mie theory \citep[][]{derooijvanderstap1984} and the refractive indices of the given materials. Studies of Solar System planets \citep[e.g.,][]{schmid2011, mclean2017} suggest that this is a good first-order approach that can improve computational runtime and efficiency for models. We acknowledge, however, that while Mie theory provides excellent fits for the spherical droplets of our liquid water clouds, water ice crystals and hydrocarbon aggregates are non-spherical in nature and can produce different optical properties \citep[e.g.,][]{heymsfield1984, bar1988}. We discuss these differences in more detail and present comparisons to models using non-spherical scattering methods for the ice clouds and hazes in Section~\ref{sec: MievsFrac}.

For both the liquid and ice water clouds, the cloud particles are distributed in size using the standard two-parameter gamma distribution of \citet[][]{hansenhovenier1974}, which is described by a particle effective radius r$_{eff}$ in $\mu$m and a dimensionless effective variance u$_{eff}$. 
Our liquid water droplets have r$_{eff}$ = 6 $\mu$m and u$_{eff}$ = 0.4 \citep[see, e.g.,][]{vandiedenhoven2007, gordon2023}. Water ice particles in Cirrus clouds display a large size range with a mean particle radius of $\sim$100 $\mu$m \citep[e.g.,][]{lohmann2016}. However, larger cloud particles are more difficult for our code to handle, so for computational efficiency, our water ice particles have r$_{eff}$ = 10 $\mu$m and u$_{eff}$ = 0.1. 
For the liquid and ice water clouds we adopted the wavelength-dependent complex refractive indices from \citet[][]{hale1973} and \citet[][]{warren2008}, respectively.

For the haze particles we calculated the optical properties of the haze at each of the six individual particle size modes and used the wavelength-dependent complex refractive indices of hydrocarbon aerosols from \citet[][]{khare1984}. We acknowledge that these optical properties were measured for Titan simulant hazes and therefore might not be a true representation of the hazes produced in the Archean eon due to the different atmospheric compositions between Titan and Earth. However, to our knowledge only one study \citep[][]{hasenkopf2010} has experimentally measured the optical properties of an Archean-analog haze, and the haze properties of \citet[][]{khare1984} produce a reasonable match to the measurement of \citet[][]{hasenkopf2010}. We therefore feel justified utilizing the \citet[][]{khare1984} optical constants for our haze models. For further discussions and comparisons between optical properties of hydrocarbon hazes across the literature, we refer the reader to \citet[][]{arney2016}.

\subsection{Model Surfaces} \label{sec: Surfs}

\begin{figure}[ht!]
    \centering
    \includegraphics[width=\linewidth]{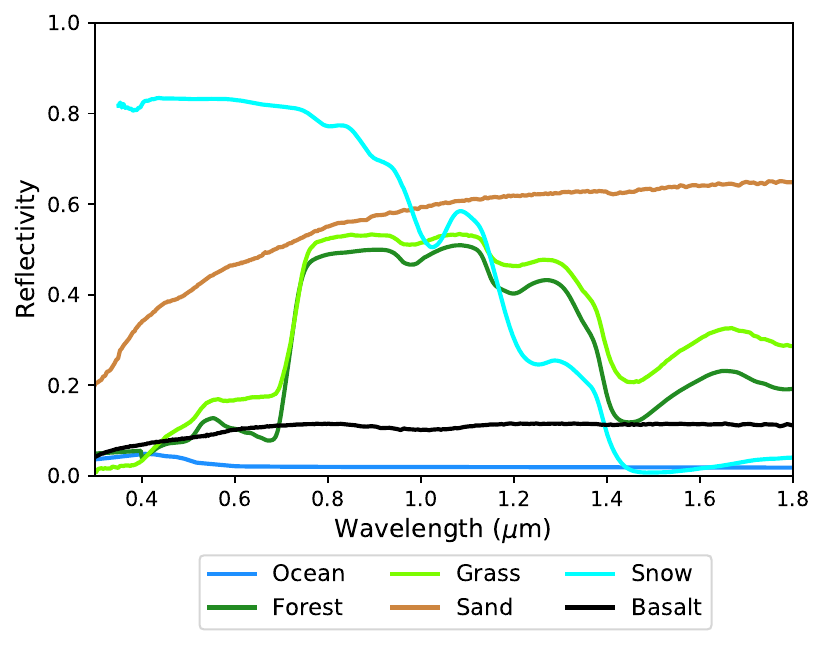}
    \caption{The six main surface albedos used in our Earth Through Time models, taken from the NASA JPL EcoStress Spectral Library and the USGS Spectral Library. Note the strong VRE in the grass and forest surfaces.}
    \label{fig:surfsmain}
\end{figure}

Our planet surfaces are modeled as ideal, depolarizing Lambertian surfaces with wavelength-dependent albedos. 
This is a common approximation in Earth modeling and retrievals \citep[see, e.g.,][and references therein]{tilstra2021}, even though for many surfaces of interest for habitability studies, like vegetation, it can lead to errors in the retrieved properties \citep[e.g.,][]{lorente2018}.
However, the disk-integrated signals of exoplanets may result in a smearing of the surface reflectance, so Lambertian surfaces are expected to provide sufficient approximations \citep[see, e.g.,][]{stam2008, kopparla2018, gordon2023}.
Nevertheless, we acknowledge that to model surfaces more accurately, bidirectional reflection (or polarization) distribution functions (BRDF/BPDF), which take into account changes in the reflected flux (or polarization) as a function of illumination and viewing angles, are needed. For example, ocean surfaces can often be wavy and rough, displaying specular features, and are best modeled as Fresnel surfaces with whitecaps \citep[see, e.g.,][and references therein]{treesstam2019, vaughan2023} using approximations such as those of \citet[][]{cox1954}. The effects of these rough ocean surfaces on their resulting signatures, such as the ocean glint, can be significant \citep[e.g.,][]{trees2022}.
Modeling BPDF surfaces for vegetation and Fresnel-reflecting surfaces for oceans is part of future work but is outside the scope of this work.
We therefore stick with Lambertian surfaces for the models presented in this study.

\begin{figure}[ht!]
    \centering
    \includegraphics[width=\linewidth]{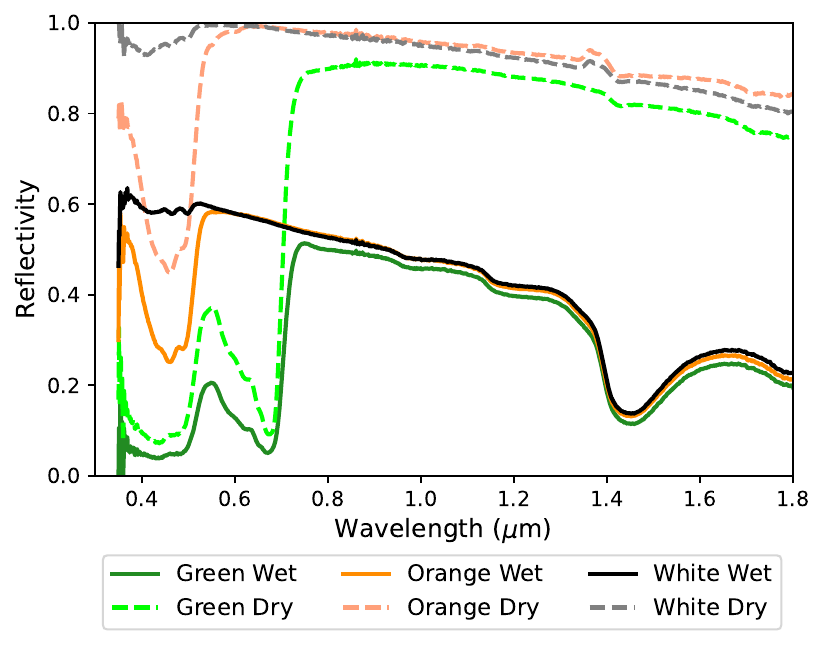}
    \caption{The surface albedos of the three biological species chosen from The Color Catalogue of Life in Ice \citep[][]{coelho2022} for our Earth Through Time models. Solid lines correspond to the fresh samples directly after collection, and the dashed lines to the same samples after they have dried.}
    \label{fig:surfsKalt}
\end{figure}

\begin{figure}[ht!]
    \centering
    \includegraphics[width=\linewidth]{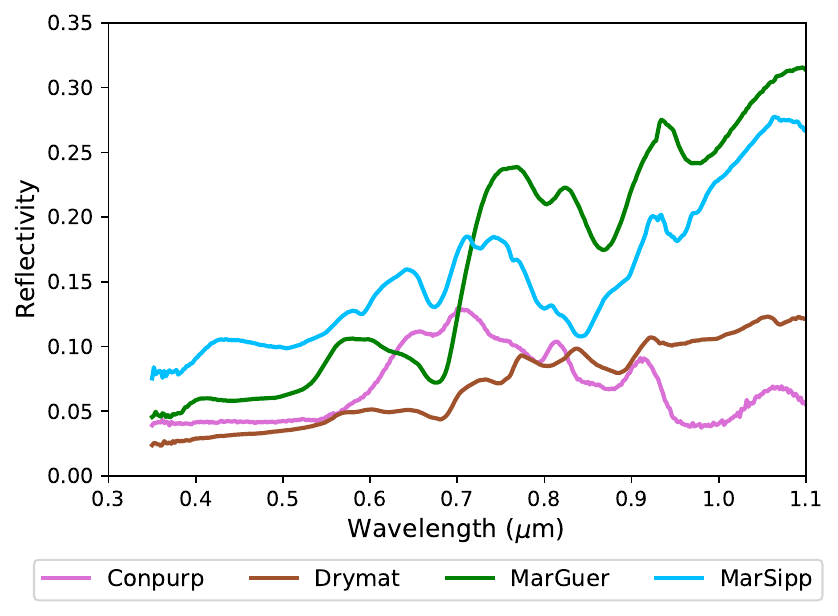}
    \caption{The four microbial surface albedos described by \citet[][]{sparks2021}. These spectra were only observed over a wavelength range of $\lambda$ = 0.35 - 1.1 $\mu$m.}
    \label{fig:surfsNiki}
\end{figure}

Here we modeled seven different categories of surfaces: ocean, forest (a combination of deciduous and conifer), grass, sand, melting snow, fresh basalt, and microbial surfaces. The reflection properties of the first six surfaces were taken from the NASA JPL EcoStress Spectral Library\footnote{\url{https://speclib.jpl.nasa.gov}} \citep[][]{baldridge2009, meerdink2019} as well as the USGS Spectral Library\footnote{\url{https://crustal.usgs.gov/speclab/QueryAll07a.php}} \citep[][]{kokaly2017} and can be seen in Fig.~\ref{fig:surfsmain}.

For the microbial surfaces we used spectra from two different studies. For the majority of our models, including all models in Section~\ref{sec: Results}, we utilized six reflection spectra from microorganisms found in the ``Color Catalogue of Life in Ice"\footnote{\url{https://zenodo.org/record/5779493}} \citep[][]{coelho2022}. These microorganisms were organized into five groups based on the colors of their pigments: orange, yellow, green, pink, or white. Additionally, their reflection spectra were measured for both fresh (i.e., wet) samples directly after collection and then for the same samples $\sim$1 week after collection (i.e., dry). Here, we utilized one wet and one dry sample from the green, orange, and white pigment groups; specifically, the samples used correspond to Chlorophyta algae (green), \textit{Brevundimonas sp.} (orange), and \textit{Bacillus sp.} (white). The reflectance spectra for these six samples are shown in Fig.~\ref{fig:surfsKalt}. Note the deep hydration feature around 1.4 $\mu$m in each of the wet samples that disappears in their corresponding dry samples. For more details regarding these samples, we refer the reader to \citet[][]{coelho2022}.

The second group of microbial surfaces we used were collected from various sites by one of the authors of this paper (M. N. Parenteau) and maintained at NASA Ames Research Center. These spectra represent environmental anoxygenic and oxygenic photosynthetic microbial mats.
We used these mat samples to represent microbial surfaces that arose during the Archean and Proterozoic eons (see Sections~\ref{sec: ArEar} and \ref{sec: ProEar}). In particular, for microbial mats on emerging continents during these time periods we used spectra of a dry \textit{Phormidium} mat (hereafter, Drymat) and a planktonic purple sulfur pool (hereafter, Conpurp). For microbial mats in shallow seas and coastal areas during these time periods we used spectra of a hypersaline mat from Guerrero Negro (hereafter, MarGuer) and a cyanobacteria mat from the Great Sippewissett Salt Marsh (hereafter, MarSipp). The reflectance spectra for these four samples are shown in Fig.~\ref{fig:surfsNiki}. For more information about these samples, we refer the reader to \citet[][]{sparks2021}, who measured the reflection and circular polarization spectra for all of these cultures and mats. 
In Section~\ref{sec: BioSurfs} we compare model spectra using these surfaces against those of \citet[][]{coelho2022}.

\subsection{Our Chosen Epochs} \label{sec: TimeExpl}

\begin{figure*}[ht!]
    \centering
    \includegraphics[width=18cm]{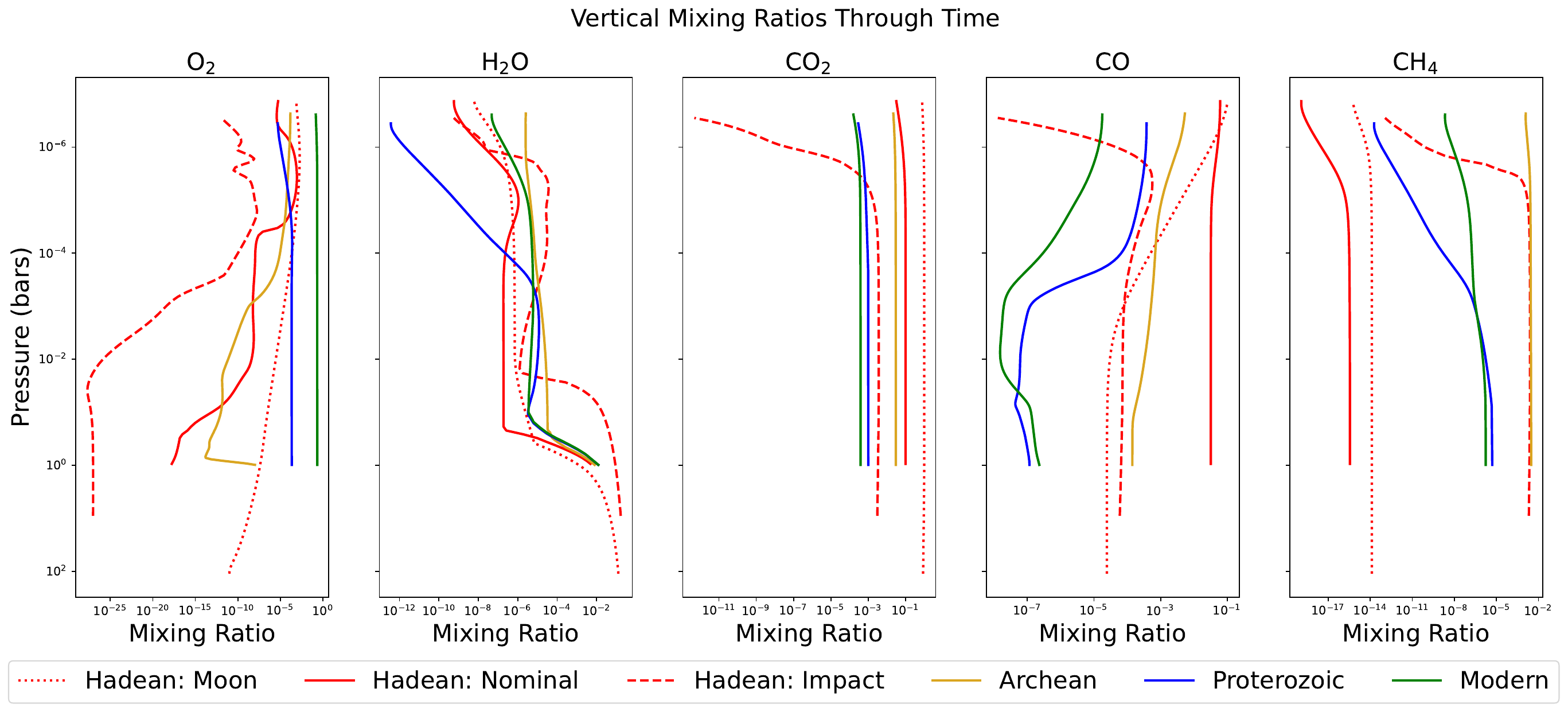}
    \caption{The change in the vertical mixing ratios for O$_2$, H$_2$O, CO$_2$, CO, and CH$_4$ over the evolution of Earth.}
    \label{fig:VMRs}
\end{figure*}

\begin{figure}[ht!]
    \centering
    \includegraphics[width=\linewidth]{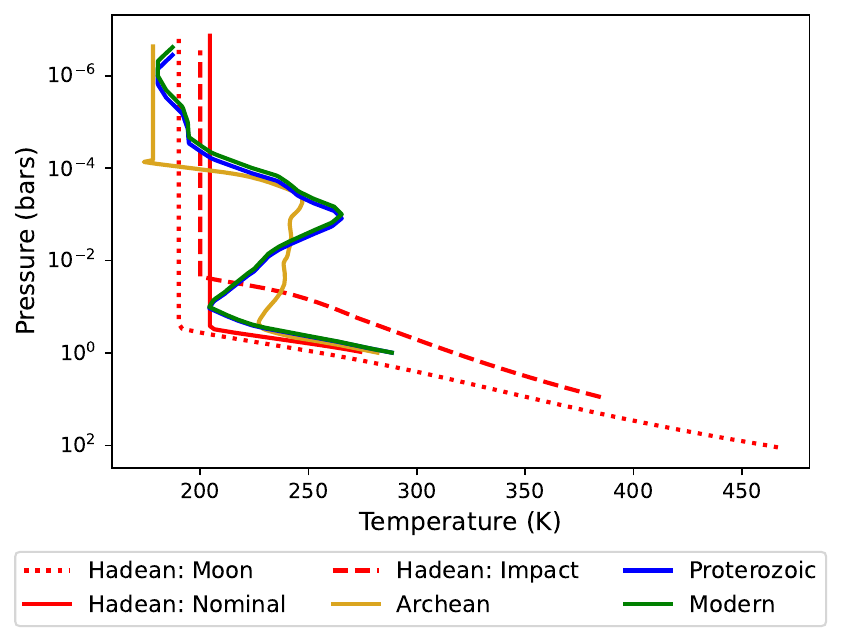}
    \caption{T-P profiles for our six Earth Through Time epochs. The surface pressure was much higher for the two impact-driven atmospheres of the Hadean eon (Hadean: Moon and Hadean: Impact), and the rise in atmospheric O$_2$ and the subsequent ozone layer led to a stronger temperature inversion in the later stages of Earth's history (Proterozoic and Modern).}
    \label{fig:TPs}
\end{figure}

To capture the evolution of Earth through geological time we used inputs from multiple studies. The vertical mixing ratios (VMRs) for the five most spectroscopically significant molecules included in all of our modeled atmospheres are shown in Fig.~\ref{fig:VMRs}. The T-P profiles for our six epochs are displayed in Fig.~\ref{fig:TPs}. Finally, Table~\ref{table:epochs} highlights the key atmospheric and surface properties for each case. For the stellar source for our calculations, we used the wavelength-dependent solar evolution correction developed by \citet[][]{claire2012} to scale the solar constant to the proper age for each epoch. These ages are also included in Table~\ref{table:epochs}.

\begin{table*}[t]
\small
\centering
\caption{Summary of key atmospheric and surface properties for our six Earth Through Time epochs. All surfaces are modeled as depolarizing Lambertian surfaces with wavelength-dependent albedos (see Figs.~\ref{fig:surfsmain} - \ref{fig:surfsNiki}).}
\begin{tabular}{c c c c c}
\hline
\hline
Epoch & Age (Ga) & Main Atmospheric Gases & Surface & $A_G$ ($\lambda$ = 0.5 $\mu$m, 0.9 $\mu$m)$^{a}$ \\
\hline

 Hadean: Moon &  4.45 &  CO$_2$, N$_2$, CO & 95\% ocean, 5\% basalt & 2.88, 0.26  \\
 Hadean: Nominal & 4.2 & N$_2$, CO$_2$, CO, H$_2$ & 80\% ocean, 20\% basalt & 0.43, 0.13 \\
 Hadean: Impact & 4.0 & H$_2$, N$_2$, CO$_2$, CH$_4$ & 70\% ocean, 30\% basalt & 2.02, 0.02 \\
 Archean & 2.7 & N$_2$, CO$_2$, CH$_4$ &  90\% ocean, 8\% Arch. land$^{b}$, & 0.47, 0.17 \\ & & & 2\% early coast$^{c}$ \\
 Proterozoic & 2.5 & N$_2$, CO$_2$, O$_2$ &  85\% ocean, 12\% Proto.~land$^{d}$, & 0.52, 0.25 \\ & & & 3\% early coast \\
 Modern & 0 & N$_2$, O$_2$, CO$_2$, H$_2$O &  70\% ocean, 8\% forest, 8\% grass, & 0.59, 0.47 \\ & & & 5\% snow, 4\% sand, 4\% basalt, \\ & & & 1\% bio$^{e}$  \\

\hline
\end{tabular}
\label{table:epochs}
\tablenotetext{a}{The geometric albedos here are for the planets with clear (i.e., cloud- and haze-free) atmospheres. Readers interested in comparing $A_G$ of all models can use Eq.~\ref{eq:normalizedflux} (setting $\alpha$ = 0$\degree$) and our models in Zenodo.}
\tablenotetext{b}{``Arch. land" represents the exposed surface of Kenorland \citep[e.g.,][]{williams1991, bindeman2018}, which we modeled as a mixture of 60\% basalt, 35\% sand, and 5\% even mixture of orange and white pigments (or continental microbial mats).}
\tablenotetext{c}{``early coast" represents a mixture of 50\% sand and 50\% even mixture of green pigments (or marine intertidal microbial mats).}
\tablenotetext{d}{``Proto.~land" represents the exposed surfaces of Kenorland \citep[e.g.,][]{williams1991, bindeman2018} after its initial breakup, as well as a few additional cratons and oceanic islands, which we modeled as a mixture of 50\% basalt, 35\% sand, 10\% snow, and 5\% even mixture of orange and white pigments (or continental microbial mats).}
\tablenotetext{e}{``bio" represents an even mixture of all pigments (or all microbial mats).}
\end{table*}

\subsubsection{Hadean Earth after Moon Forming Impact} \label{sec: HadMoon}

Shortly after the formation of the Earth around 4.54 Ga, it is thought that a Mars-sized proto-planet named Theia collided with the proto-Earth, with the resulting ejecta forming the Moon \citep[e.g.,][]{hartmann1975, canup2001}. This moon-forming giant impact would have melted the entire proto-Earth mantle down to the core, resulting in a global magma ocean. All volatile elements (including, e.g., H$_2$O, CO$_2$, etc.) would have then partitioned between the magma and the atmosphere according to their solubility. Since water is much more soluble in magma than CO$_2$ \citep[e.g.,][]{blank1994, gardner1999}, most of the water from proto-Earth would have been contained in the magma ocean, while most of the total reservoir of CO$_2$ (about 100 bar) would be in the atmosphere \citep[e.g.,][]{zahnle2010}. The magma would solidify in $\sim$1-2 million years, followed by a rapid condensation of the ocean, leaving behind a warm $\sim$500 K planet with 100 bars of CO$_2$ and a shallow liquid ocean \citep[e.g.,][]{zahnle2007}. Indeed, the oldest found zircon crystals are believed to be as old as $\sim$4.4 Ga \citep[e.g.,][]{wilde2001}, suggesting that liquid water formed on proto-Earth rather early. Although continental crust was most likely absent, high levels of activity from mantle plumes probably created multiple exposed ocean islands \citep[see, e.g.,][and references therein]{korenaga2021}. The CO$_2$ in the atmosphere probably then reacted with the exposed seafloor during the planetary cooling process and subducted back into the planet, thereby removing the thick CO$_2$ atmosphere in 20-100 million years\citep[e.g.,][]{zahnle2010}.

Our first Hadean Earth model (hereafter, Hadean: Moon) is a simulation of the Earth when the atmosphere was CO$_2$-dominated for 20-100 million years after the moon-forming giant impact. This atmosphere also contained large amounts of N$_2$ and H$_2$O as well as trace amounts of CO, H$_2$, CH$_4$, and O$_2$. The surface was dominated by a shallow ocean with about 5\% of the globe also covered by basaltic exposed islands.

\subsubsection{Nominal Hadean Earth} \label{sec: HadNom}

For the majority of the Hadean eon ($\sim$4.54 - 4.03 Ga), the early Earth potentially had an N$_2$-CO$_2$ dominated atmosphere with trace amounts of volcanic and photochemical H$_2$ and CO \citep[e.g.,][]{catling2020}. Additionally, large amounts of continental crust could have developed as the Earth continued to cool from the moon-forming giant impact \citep[e.g.,][]{guo2020}. The growth of the continents could have outpaced the condensation of the still-shallow ocean, and ocean islands from mantle plumes continued to provide more exposed land, such that approximately 20\% or more of the Earth’s surface could have been covered by exposed continental crust \citep[e.g.,][]{korenaga2021}.

Our second Hadean Earth model (hereafter, Hadean: Nominal) is a simulation of the Earth during the mid-Hadean when it had a 1 bar N$_2$-dominated atmosphere with 10\% CO$_2$ and trace amounts of O$_2$ and H$_2$O. Our photochemical simulations also assumed that CO, H$_2$, SO$_2$, and H$_2$S were emitted to the atmosphere at rates calculated in \citet[][]{wogan2020}. As with the Hadean: Moon model, the surface of our Hadean: Nominal model was dominated by a shallow ocean but now with 20\% of the Earth covered by exposed basaltic crust.

\subsubsection{Hadean Earth after Asteroid Impact} \label{sec: HadImp}

Throughout the Hadean eon Earth was struck by asteroid impactors. The largest of these impactors possessed cores that would have delivered metallic iron that could potentially reduce the atmosphere and vaporize the ocean, even potentially melting the crust down a few tens of kilometers \citep[e.g.,][]{zahnle2010, zahnle2020}. This would have created a transient H$_2$-CH$_4$-N$_2$-CO$_2$ atmosphere that more closely resembles Neptune’s than modern Earth’s. Additionally, a large impact combined with a relatively dry early-Earth stratosphere could allow for the buildup of thick photochemical hazes. The transient, hazy atmosphere would have lasted for millions of years until the hydrogen escaped to space and the skies cleared \citep[e.g.,][]{zahnle2020, wogan2023}.

Our third Hadean Earth model (hereafter, Hadean: Impact) is a snapshot of the Hadean atmosphere towards the end of the eon, $\sim$15,000 years after a Ceres-sized ($\sim$900 km) asteroid impact. We modeled a $\sim$9 bars, H$_2$-dominated atmosphere with large amounts of CH$_4$, N$_2$, and CO$_2$ that was rich in hydrocarbon haze. Our photochemical simulations also assumed trace amounts of O$_2$, H$_2$O, CO, HCN, and NH$_3$. As with our two previous Hadean Earth models, the surface of our Hadean: Impact model was dominated by an ocean but now with 30\% of the Earth covered by basaltic crust that was exposed due to the impact.

\subsubsection{Archean Earth} \label{sec: ArEar}

The Archean eon began $\sim$4.03 Ga and lasted until the first hypothesized Snowball Earth during the Huronian glaciation event, when our atmosphere began transitioning from anoxic to oxic with the first substantial rise in O$_2$ \citep[$\sim$2.5 Ga; e.g.,][]{tang2013, young2019}.
From the early to mid-Archean, the Earth was mostly a water world \citep[e.g.,][]{korenaga2021}. Towards the end of the eon, increased mantle convection and tectonic activity is believed to have caused several cratons to rise above the ocean surface and combine into the first supercontinents \citep[e.g.,][]{deKock2009, rogers1996, bindeman2018, williams1991}.
It is also assumed that the Archean was the first geologic eon with a prevalent microbial biosphere \citep[e.g.,][]{dodd2017, djokic2017, allwood2006}, although the precise timing of the emergence of the first life on Earth is not known. 
The atmosphere during this eon is thought to have had very low levels of O$_2$, which could have supported the formation of an organic haze generated by CH$_4$ photolysis \citep[e.g.,][]{claire2014}.

Our Archean Earth model is a simulation of the Earth after the evolution of oxygenic photosynthesis but prior to any substantial increase in O$_2$ in the atmosphere. We used the Archean Earth atmospheric model from \citet[][]{arney2016}, which assumed a weakly reducing atmosphere dominated by N$_2$ and CO$_2$ with large amounts of CH$_4$ and CO. This atmosphere also included H$_2$O and H$_2$ as well as small amounts of atmospheric O$_2$ and NO. As in \citet[][]{arney2016}, our Archean atmosphere was rich in hydrocarbon haze similar to modern-day Titan. The surface of our Archean model was dominated by a deep ocean and included a single supercontinent covering $\sim$10\% of the globe \citep[][]{bindeman2018}. 
2\% of this land we modeled as coastal regions with 50\% sand and 50\% microorganisms, and the remaining 8\% was a combination of basalt, sand, and microorganisms.

\subsubsection{Proterozoic Earth} \label{sec: ProEar}

The Proterozoic eon was the longest in Earth's history, lasting from the first rise in O$_2$ concentration in the planet's atmosphere and oceans ($\sim$2.5 Ga) to just before the appearance of diverse complex life ($\sim$0.5388 Ga). Multiple geological and ecological changes occurred throughout this eon, including the transition from a reducing to an O$_2$-rich atmosphere through the Great Oxidation Event (GOE) \citep[e.g.,][]{holland2006, lyons2014}; a series of global glaciation events \citep[e.g.,][]{tang2013}; the rise and breakup of multiple new supercontinents \citep[e.g.,][]{rogers2002, li2008, nance2019}; and 
the rise of eukaryotic organisms and multicellular life \citep[e.g.,][]{schirrmeister2013}.

Our Proterozoic Earth model is a simulation of the Earth at the beginning of the eon, before the start of the GOE but after the introduction of more O$_2$ in the atmosphere from cyanobacteria. We utilized the Proterozoic Earth atmospheric model from \citet[][]{arney2016}, which assumed an N$_2$-dominated atmosphere still with large levels of CO$_2$ and atmospheric H$_2$O but with 0.1\% the present atmospheric level of O$_2$, consistent with the findings in \citet[][]{planavsky2014}. These photochemical simulations also predicted trace amounts of CH$_4$, H$_2$, CO, and O$_3$. The surface of our Proterozoic model was dominated by a deep ocean with 15\% of the globe covered by land. 3\% of this land was modeled as coastal regions with 50\% sand and 50\% microorganisms, and the remaining 12\% was a combination of basalt, sand, microorganisms, and snow.

\subsubsection{Phanerozoic (Modern) Earth} \label{sec: ModEar}

As Earth's current geologic eon, the Phanerozoic eon began with the Cambrian explosion ($\sim$0.5388 Ga) \citep[e.g.,][]{marshall2006} that marked the rapid proliferation and diversification of complex life.
Effects of this complex life, specifically those from vegetation, on the planet's climate and resulting flux can be seen in observations of earthshine \citep[e.g.,][]{woolf2002, palle2009}. Increased vegetation also led to the last major rise in atmospheric O$_2$ and the strengthening of the ozone layer \citep[e.g.,][]{kasting1980, segura2003}.
Tectonic forces collected the existing landmasses into the most recent supercontinent Pangaea \citep[e.g.,][]{dietz1970}, which then separated into the current continents.

Our Modern Earth model is a simulation of the Earth as it appears today. The atmosphere is N$_2$-dominated with 21\% O$_2$ and $\sim$0.0366\% CO$_2$, followed by present-day trace amounts of CO, CH$_4$, O$_3$, N$_2$O, and NO. The surface of our Modern Earth is covered by 70\% deep oceans, 5\% snow, and 25\% land that is dominated by vegetation (i.e., forest and grass).

\section{Results \& Comparisons} \label{sec: Results}

\begin{figure*}[ht!]
    \centering
    \includegraphics[width=16cm]{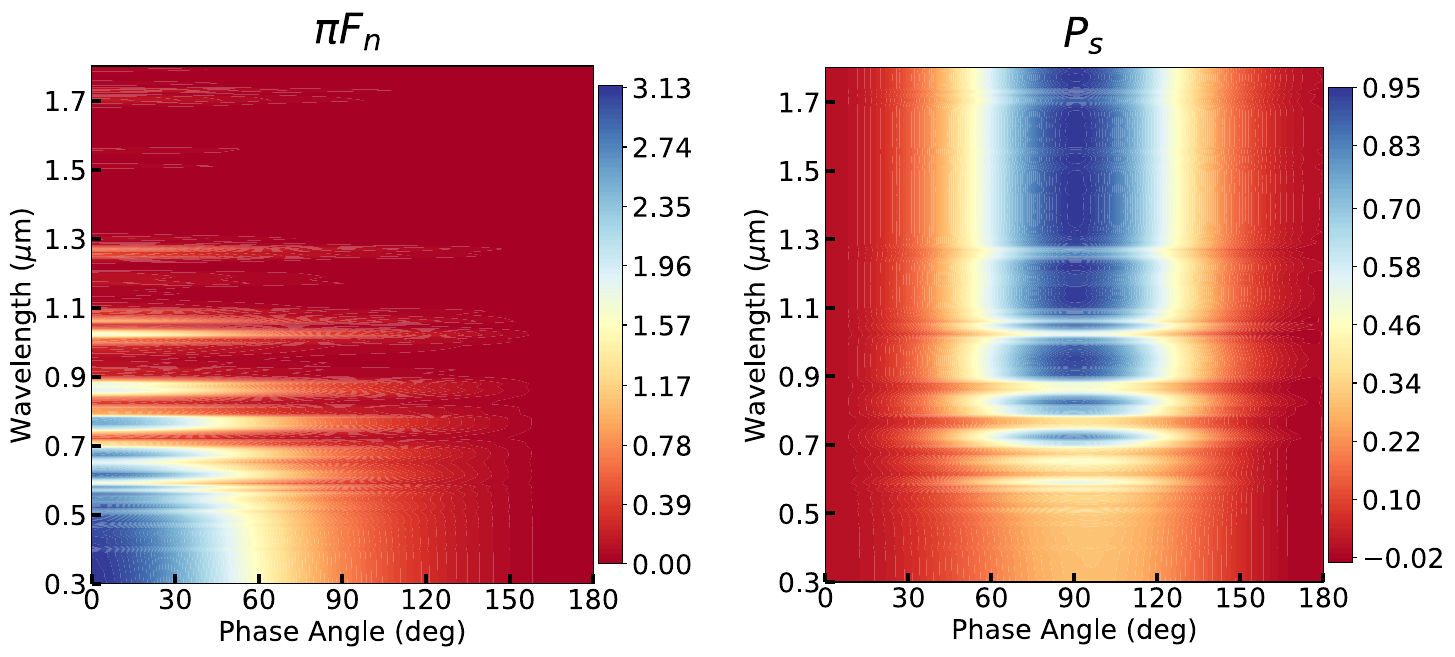}
    \caption{Reflected total normalized flux $\pi{F_{n}}$ (left panel) and signed degree of linear polarization P$_s$ (right panel) as functions of the planetary phase angle $\alpha$ and wavelength $\lambda$ for our Hadean: Moon model. Note the differences in the colorbars for the two plots.}
    \label{fig:HadMooncontour}
\end{figure*}

\begin{figure*}[ht!]
    \centering
    \includegraphics[width=16cm]{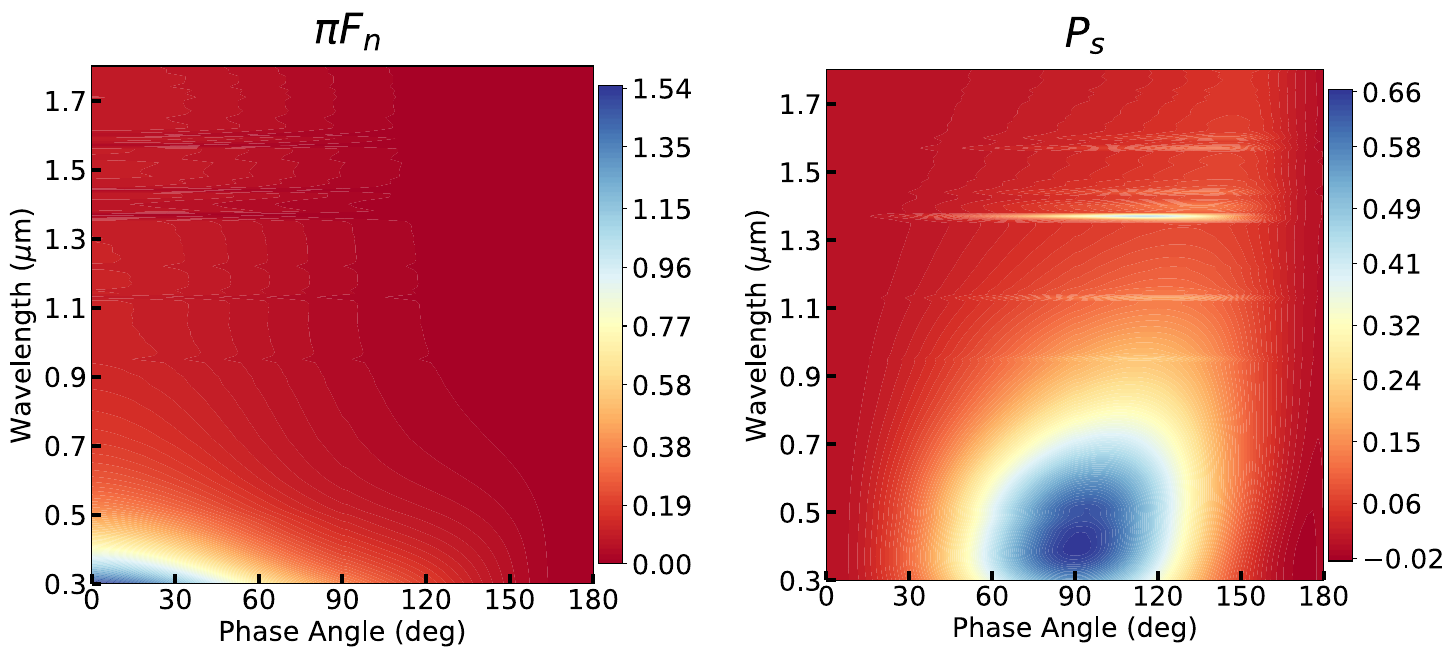}
    \caption{Same as Fig.~\ref{fig:HadMooncontour}, except for our Hadean: Nominal model.}
    \label{fig:HadNomcontour}
\end{figure*}

\begin{figure*}[ht!]
    \centering
    \includegraphics[width=16cm]{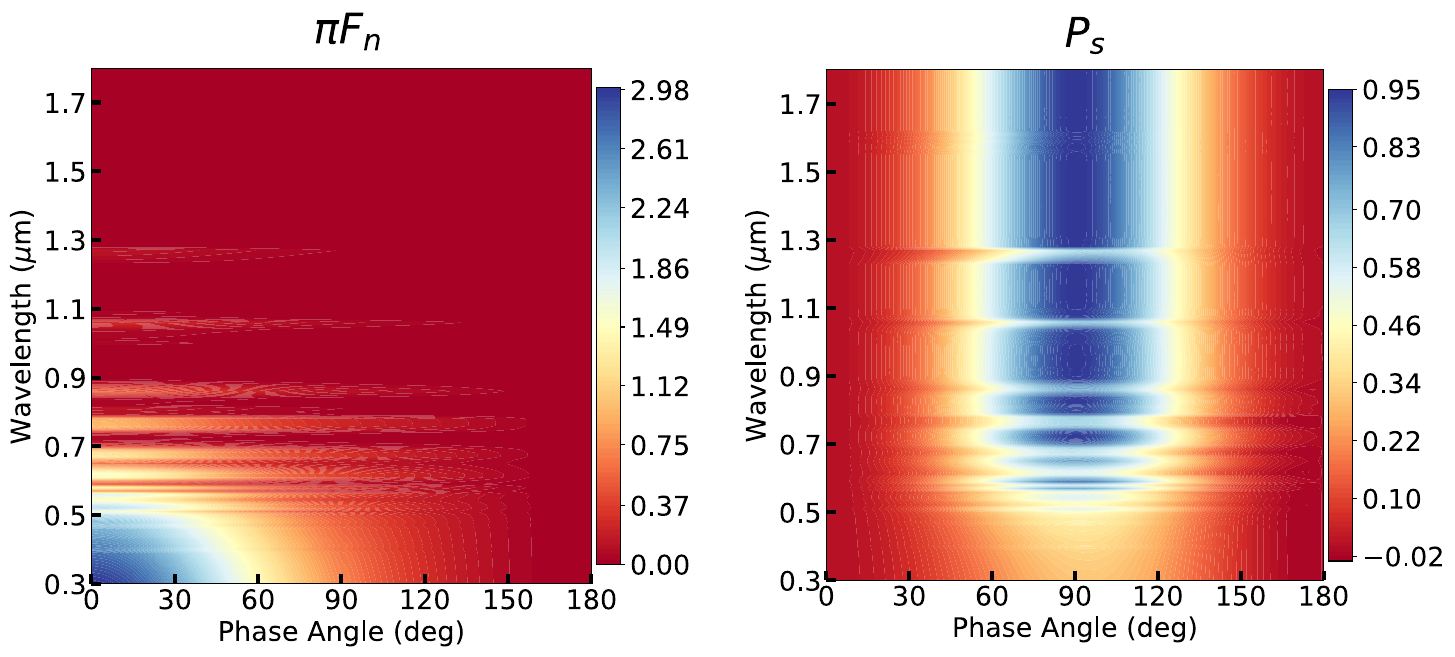}
    \caption{Same as Fig.~\ref{fig:HadMooncontour}, except for our Hadean: Impact model with a clear atmosphere.}
    \label{fig:HadImpclearcontour}
\end{figure*}

\begin{figure*}[ht!]
    \centering
    \includegraphics[width=16cm]{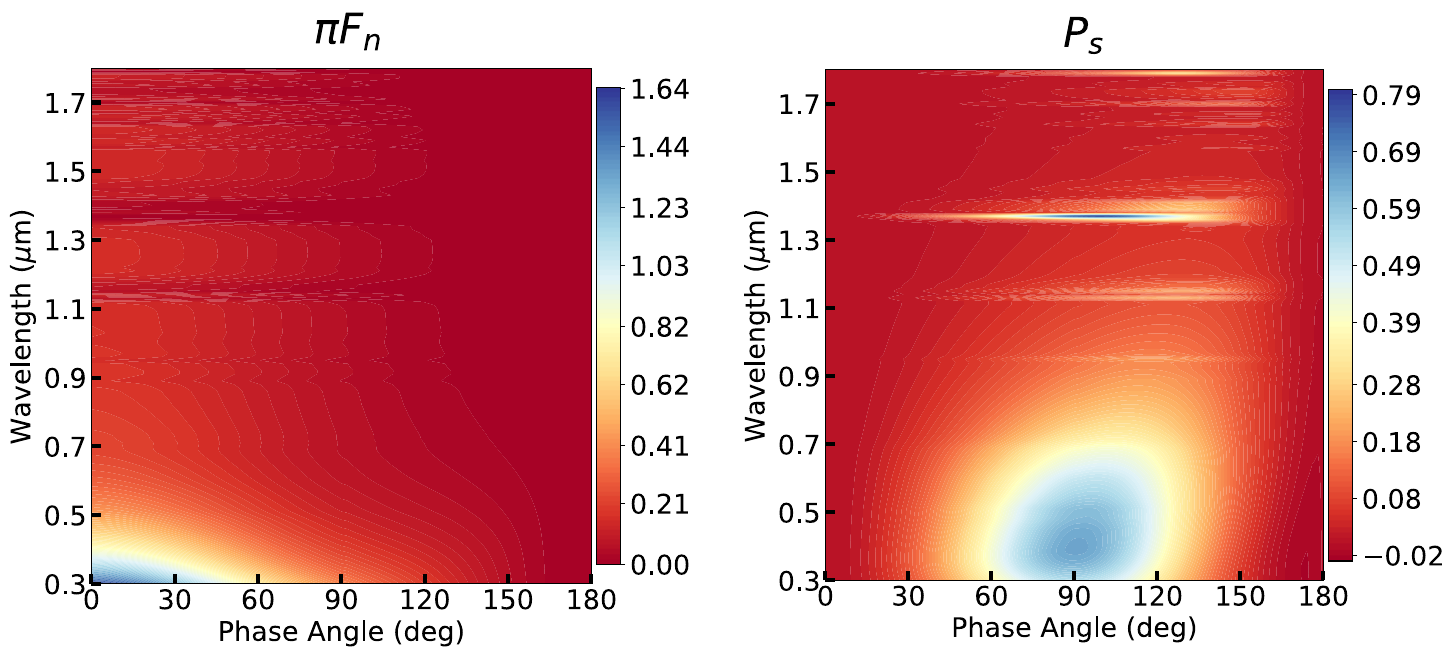}
    \caption{Same as Fig.~\ref{fig:HadMooncontour}, except for our Archean model with a clear atmosphere.}
    \label{fig:AENHclearKBiocontour}
\end{figure*}

\begin{figure*}[ht!]
    \centering
    \includegraphics[width=16cm]{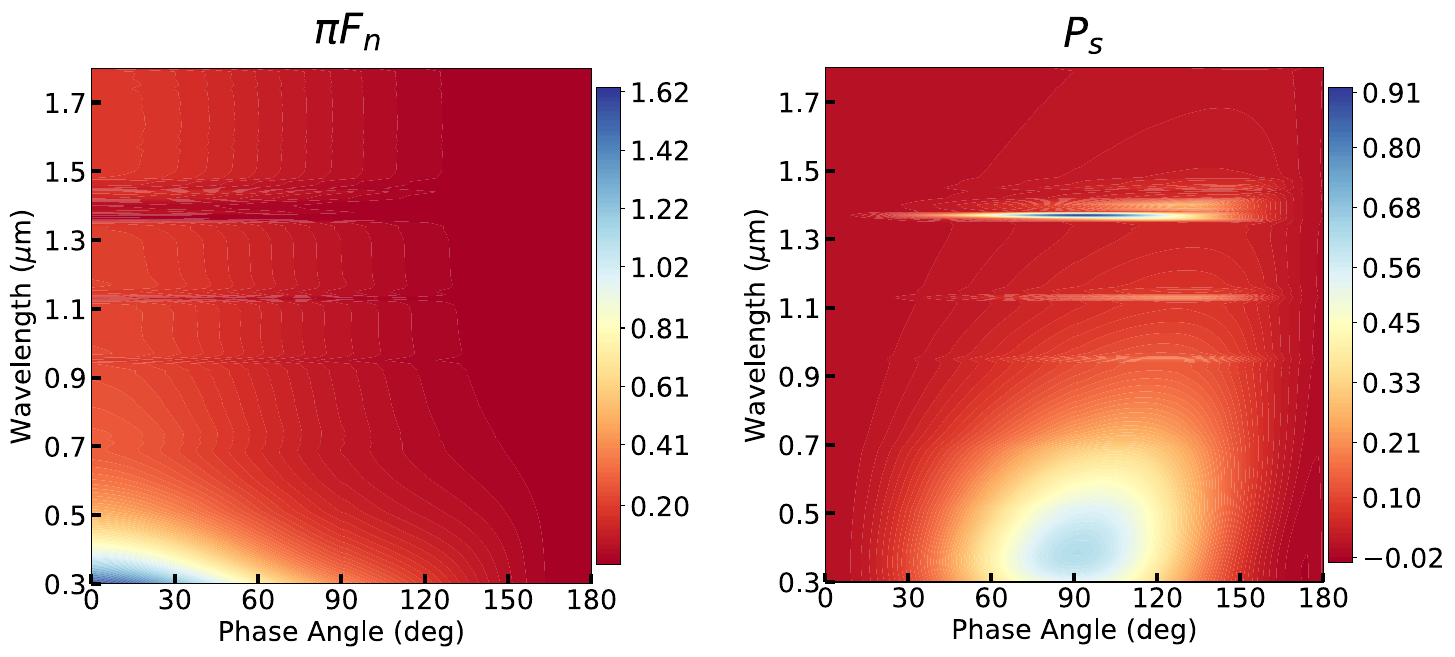}
    \caption{Same as Fig.~\ref{fig:HadMooncontour}, except for our Proterozoic model with a clear atmosphere.}
    \label{fig:PEclearKBiocontour}
\end{figure*}

\begin{figure*}[ht!]
    \centering
    \includegraphics[width=16cm]{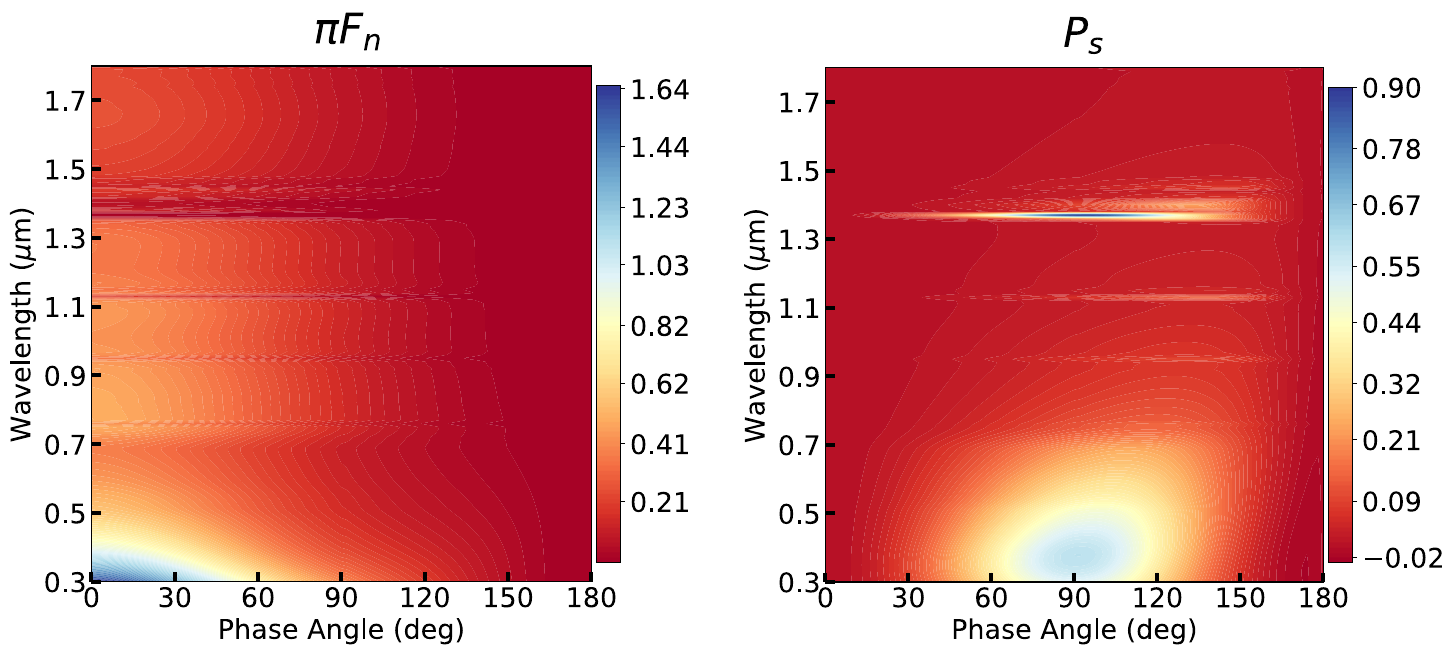}
    \caption{Same as Fig.~\ref{fig:HadMooncontour}, except for our Modern model with a clear atmosphere.}
    \label{fig:MEclearKBiocontour}
\end{figure*}

\begin{figure*}[ht!]
    \centering
    \includegraphics[width=15cm]{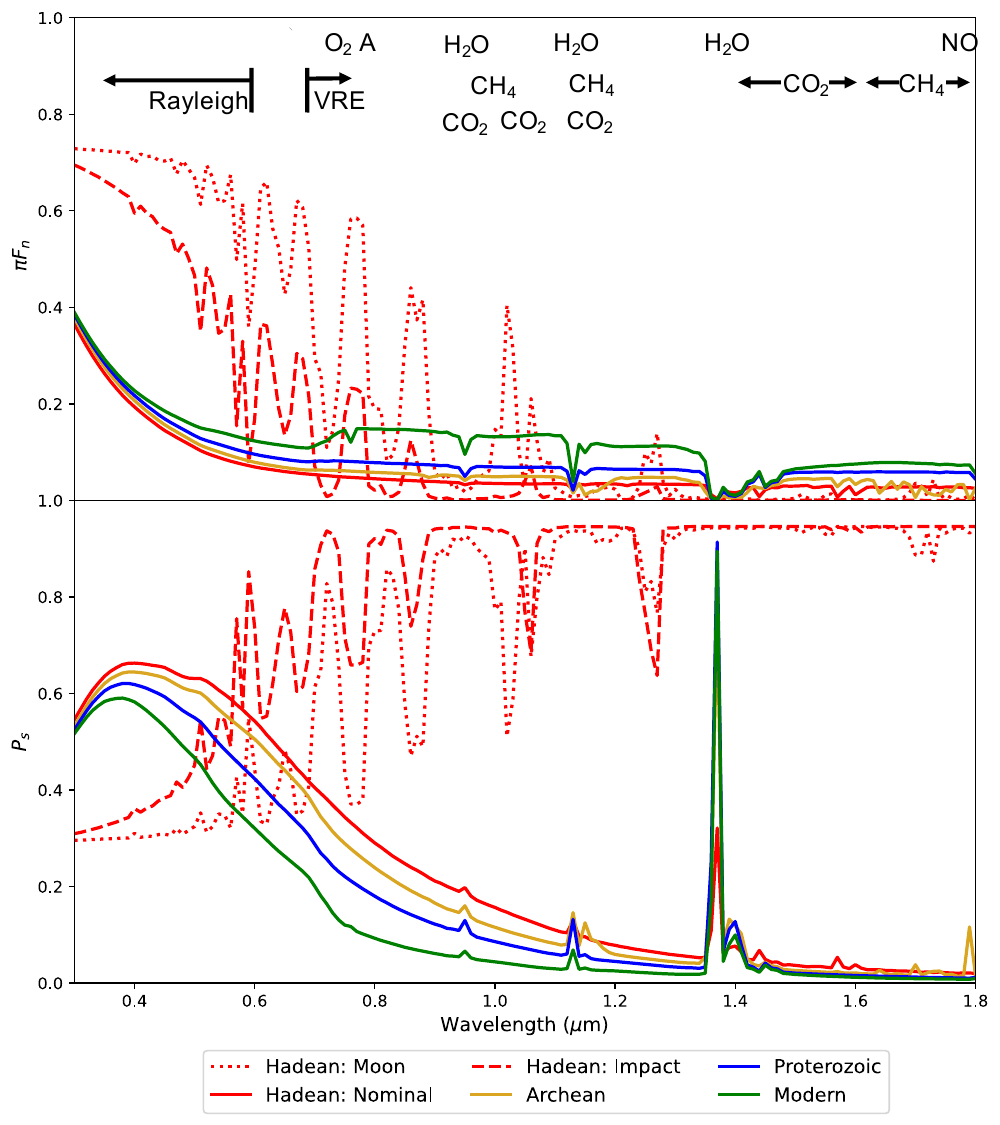}
    \caption{Cross-sections through the left (top panel) and right (bottom panel) panels of Figs.~\ref{fig:HadMooncontour} - \ref{fig:MEclearKBiocontour} at $\alpha = 90\degree$ to show the spectral differences of each of our six Earth Through Time epochs. Note the distinguishing spectral features of different species across different epochs as well as a noticeable VRE in our Modern Earth model.}
    \label{fig:clearspectra}
\end{figure*}

In this section, we present the total normalized flux and signed degree of linear polarization signatures of starlight reflected by our six Earth Through Time epochs. 
Our models have a constant $\Delta\lambda$ of 10 nm, corresponding to spectral resolving powers of R$_{VIS}$ $\sim 50$ and R$_{NIR}$ $\sim 150$.
Unless otherwise stated, our models were generated using the DAP code.

\subsection{Clear Atmospheres} \label{sec: clear}

Fig.~\ref{fig:HadMooncontour} through Fig.~\ref{fig:MEclearKBiocontour} display the total normalized flux ($\pi{F_{n}}$) and signed degree of linear polarization (P$_s$) of the reflected flux as functions of both $\lambda$ and $\alpha$ for our six epochs. All of these models possess clear (i.e., cloud- and haze-free) atmospheres, with atmospheric species and surfaces described in detail in Section~\ref{sec: TimeExpl} and highlighted in Table~\ref{table:epochs}.

P$_s$ for all six models peak around $\alpha$ = 90$\degree$ due to Rayleigh scattering and the large contribution of the dark ocean on each surface (see Fig.~\ref{fig:surfsmain} and Table~\ref{table:epochs}), as expected \citep[see, e.g.,][]{hansentravis1974, stam2008}.
Similarly, $\pi{F_{n}}$ peaks at $\alpha$ = 0$\degree$ (i.e., when the planet is fully illuminated) and decreases smoothly with phase until $\alpha$ = 180$\degree$ (i.e., when we see the planet's night side). All models also show the planets are brighter (i.e., larger $\pi{F_{n}}$) at shorter $\lambda$, where the atmospheric optical thickness is the largest, and gradually get darker with increasing $\lambda$. The hotter atmospheres and higher surface pressures (see Fig.~\ref{fig:TPs}) of the Hadean: Moon and Hadean: Impact scenarios, combined with higher concentrations of CO$_2$ and CH$_4$, respectively, in their atmospheres (see Fig.~\ref{fig:VMRs}), increased the NIR absorption for these models. This darkened the planets and led to lower $\pi{F_{n}}$ compared to our other scenarios. The increased absorption diminished the multiply scattered background light, however, resulting in higher P$_s$ for these models.

As future missions aimed at characterizing habitable worlds will be focusing on direct imaging and therefore will be optimized for viewing the planets near quadrature (i.e., $\alpha$ = 90$\degree$), Fig.~\ref{fig:clearspectra} displays $\pi{F_{n}}$($\lambda$) and P$_s$($\lambda$) for all six models at $\alpha$ = 90$\degree$. The increase in global snow coverage and vegetation in the Modern model compared to the Archean or Proterozoic models leads to higher surface reflectivity, thereby increasing the $\pi{F_{n}}$ and lowering the P$_s$ of the planet. Additionally, the VRE around 0.7 $\mu$m, due to absorption by chlorophyll in the vegetated surfaces, is apparent in both the $\pi{F_{n}}$ and P$_s$ of the Modern spectra. 
The atmospheric evolution and resulting change in VMRs of key atmospheric species (see Fig.~\ref{fig:VMRs}) leads to different features dominating the spectra over time, thereby differentiating the model Earths across the epochs.
Most noticeable are the strongly polarized NIR H$_{2}$O bands near 0.93, 1.12, and 1.35 - 1.4~$\mu$m whose depths change depending on their respective mixing ratios in their models. We can also see the strong O$_{2}$ A-band at 0.76~$\mu$m in the Modern model; the 1.15, 1.64, and 1.72~$\mu$m CH$_4$ bands and 1.78~$\mu$m NO band in the Archean model; and the 1.44~$\mu$m CO$_2$ band in the Proterozoic and Modern models.

\subsection{Effects of Clouds and Hazes} \label{sec: cloudhaze}

\begin{figure}[ht!]
    \centering
    \includegraphics[width=\linewidth]{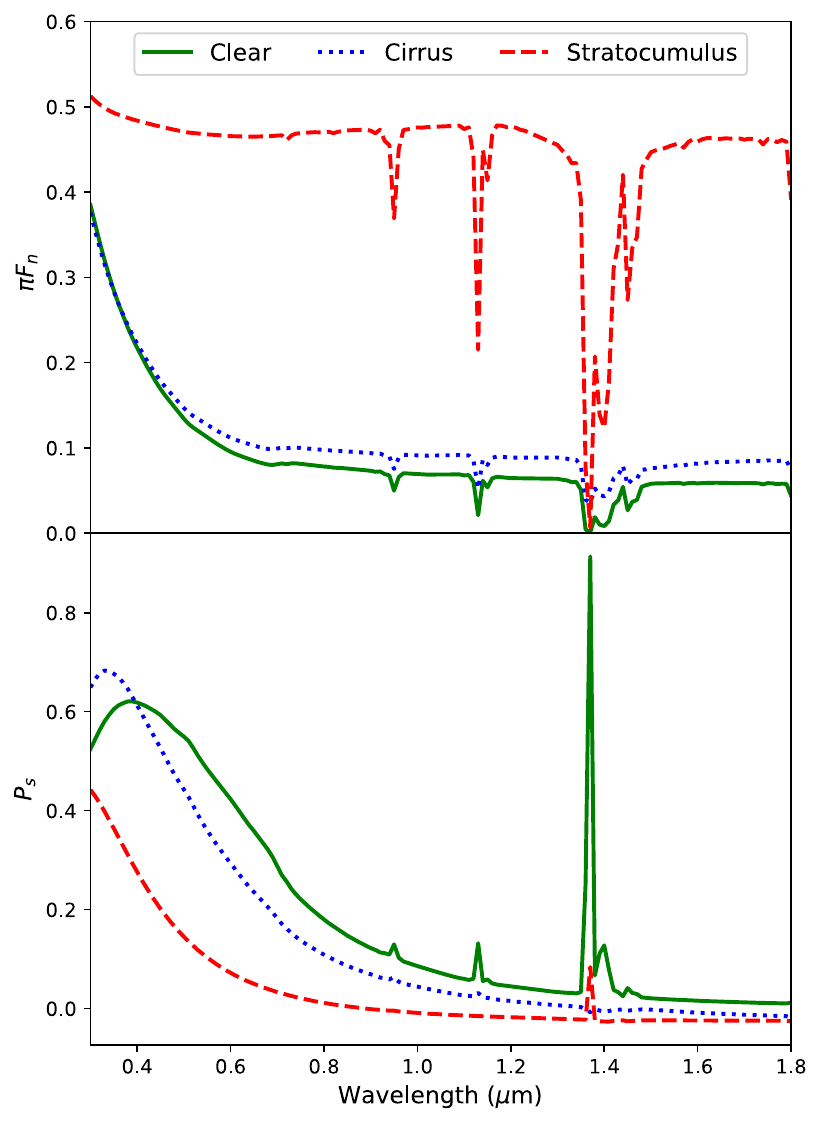}
    \caption{$\pi{F_{n}}$($\lambda$) (top) and P$_s$($\lambda$) (bottom) from our Proterozoic model for atmospheres that are: clear (solid green lines), with Cirrus clouds (dotted blue lines), or with Stratocumulus clouds (dashed red lines). All models are shown at quadrature. The addition of clouds increases the reflectivity of the planet but flattens the spectra, especially at shorter $\lambda$. The three models are more easily distinguishable in P$_s$.}
    \label{fig:ProtoClouds}
\end{figure}

Aerosols in an atmosphere play a significant role in determining the overall polarization state of the planet. Here we discuss the signatures of our models for planets that are completely covered by aerosols.
Fig.~\ref{fig:ProtoClouds} shows our Proterozoic epoch model at quadrature for three different cases: a clear atmosphere (green solid lines), an atmosphere with a layer of Cirrus clouds (dotted blue lines), and an atmosphere with a layer of Stratocumulus clouds (dashed red lines); see Section~\ref{sec: CloudsHazes} for the properties of these clouds. 
As expected, the addition of water clouds in the atmosphere reduces or even flattens some absorption and surface features, and the increased albedo of the cloud droplets and multiple scattering within the clouds lead to higher $\pi{F_{n}}$ and lower P$_s$. While the Stratocumulus clouds are more optically thick than the Cirrus clouds, the higher altitude of the latter block more light from reaching lower in the atmosphere than the former, thus further suppressing the 1.4 $\mu$m H$_2$O absorption. We can see that P$_s$ does a better job than $\pi{F_{n}}$ at distinguishing the three models. While the normalized flux spectra between the clear and Cirrus cases overlap across multiple wavelengths, with the maximum absolute difference in $\pi{F_{n}}$ being $\sim$0.035 in the 1.4 $\mu$m H$_2$O band and $\sim$0.025 in the continuum, the polarization spectra show clear separation between the models, with the maximum absolute difference in P$_s$ being $\sim$0.9 in the 1.4 $\mu$m H$_2$O band and $\sim$0.125 in the continuum.

\begin{figure}[ht!]
    \centering
    \includegraphics[width=\linewidth]{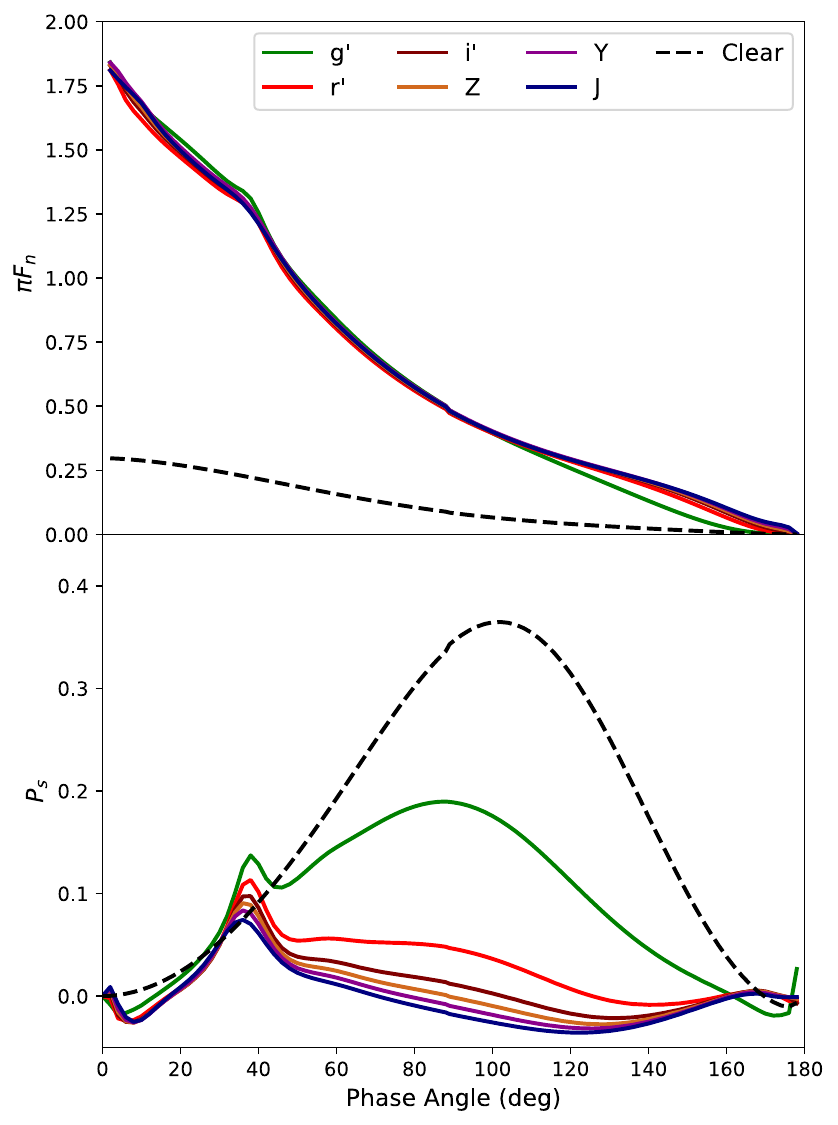}
    \caption{$\pi{F_{n}}$($\alpha$) (top) and P$_s$($\alpha$) (bottom) from our Proterozoic model with Stratocumulus clouds, shown at the center $\lambda$ of various bandpasses. For comparison, curves from our Proterozoic model with clear atmospheres are also shown for r$^{\prime}$ band (dashed black lines). Diagnostic features of the liquid water clouds (primary rainbow at $\alpha \approx$ 40$\degree$) can be seen slightly in $\pi{F_{n}}$($\alpha$) and more clearly in P$_s$($\alpha$).}
    \label{fig:CloudPhases}
\end{figure}

Fig.~\ref{fig:CloudPhases} displays phase curves of $\pi{F_{n}}$ and P$_s$ in different observational bands for our Proterozoic model with Stratocumulus clouds. Phase curves from our clear atmosphere Proterozoic model are also included at r$^{\prime}$ band for comparison (dashed black lines). As expected, the clear atmosphere model shows the smooth decrease of $\pi{F_{n}}$ towards larger $\alpha$ and the peak P$_s$ around $\alpha$ $\approx$ 90 - 100$\degree$. P$_s$ is more sensitive to the introduction of clouds than $\pi{F_{n}}$, and although the $\pi{F_{n}}$($\alpha$) for all $\lambda$s overlap and are therefore indistinguishable, the P$_s$($\alpha$) show the wavelength dependence of the rainbow feature for the liquid water clouds, with the rainbow peak shifting to smaller $\alpha$ with increasing $\lambda$ \citep[see e.g.,][]{bailey2007, karalidi2012rainbow}. Additionally, the rainbow feature becomes stronger with increasing $\lambda$, becoming a global maximum in P$_s$($\alpha$), and its strength can change depending on the cloud altitude and optical thickness \citep[for more details, see e.g.,][]{gordon2023}.

\begin{figure}[ht!]
    \centering
    \includegraphics[width=\linewidth]{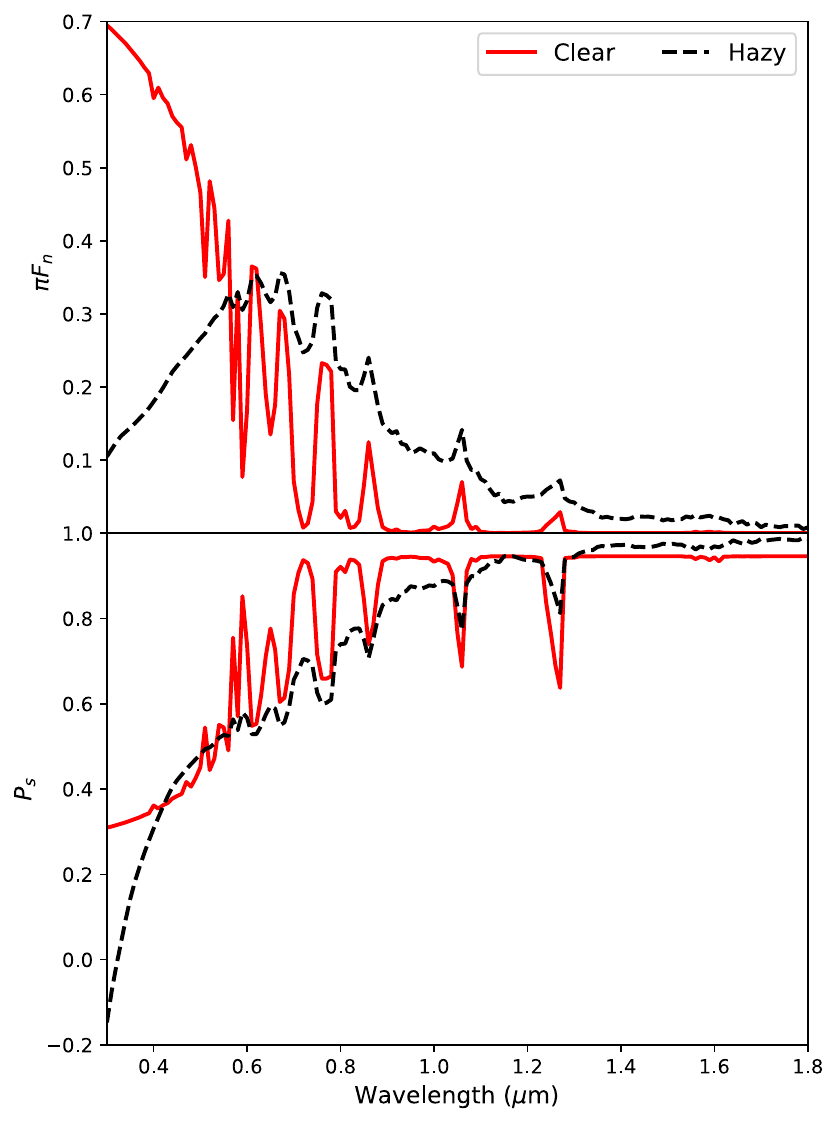}
    \caption{$\pi{F_{n}}$($\lambda$) (top) and P$_s$($\lambda$) (bottom) at $\alpha$ = 90$\degree$ for our Hadean: Impact model with either a clear (solid red lines) or hazy (dashed black lines) atmosphere. Hazes flatten the spectra and flip the polarization of the short-wavelength light.}
    \label{fig:HadImpHaze}
\end{figure}

Fig.~\ref{fig:HadImpHaze} shows the influence of hydrocarbon haze in the atmosphere of our Hadean: Impact scenario on the resulting planetary $\pi{F_{n}}$($\lambda$) and P$_s$($\lambda$) at quadrature. Due to their opaque nature the hazes block the light from reaching lower in the atmosphere, thereby flattening the spectra and reducing NIR absorption features but increasing the reflectivity of the planet at these longer wavelengths. 
As in \citet[][]{arney2016}, the hazes also result in reddening of the unpolarized light color of the planet.
The increased UV absorption combined with increased scattering within the aerosol particles decreases the $|$P$_s$$|$ of the light in the visible but increases it in the UV. The change in sign of P$_s$ to negative in the UV (i.e., the light is now polarized parallel to the scattering plane) is due to first order scattered light from the aerosol particles \citep[see, e.g.,][]{karalidi2011, karalidi2013flux}.

\begin{figure}[ht!]
    \centering
    \includegraphics[width=\linewidth]{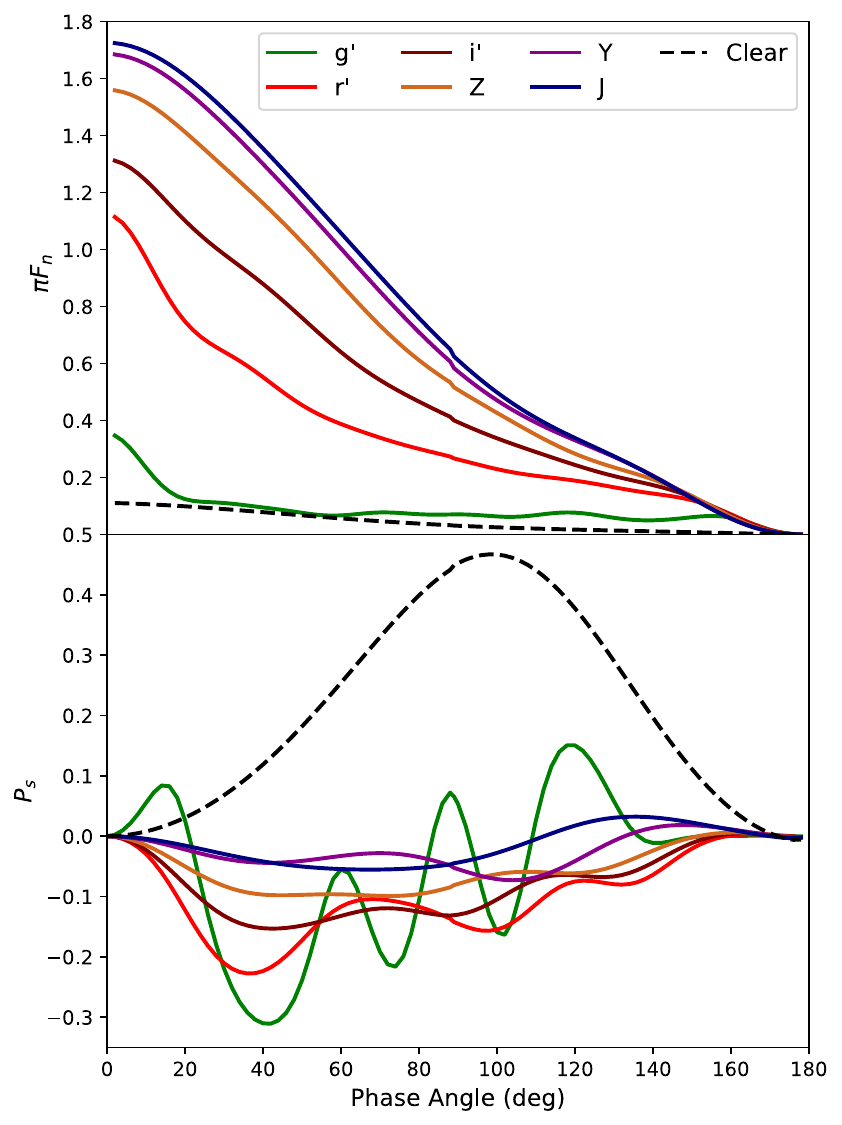}
    \caption{$\pi{F_{n}}$($\alpha$) (top) and P$_s$($\alpha$) (bottom) of a homogeneous planet with an ocean surface and Achean atmosphere containing hydrocarbon hazes, shown at the center $\lambda$ for various bandpasses. For comparison, curves of a homogeneous ocean planet with a clear Archean atmosphere are also shown for the i$^{\prime}$ band (dashed black lines). Features of the hazes are apparent in both plots but are more distinguishing in the P$_s$($\alpha$).}
    \label{fig:ArchHaze}
\end{figure}

Fig.~\ref{fig:ArchHaze} shows $\pi{F_{n}}$($\alpha$) and P$_s$($\alpha$) at different observational bands for a homogeneous ocean planet (i.e., the entire surface is now modeled as a depolarizing Lambertian ocean; see Section~\ref{sec: Surfs} and Fig.~\ref{fig:surfsmain}) with an Archean atmosphere containing hydrocarbon haze. For comparison, phase curves from a homogeneous ocean planet with a clear Archean atmosphere are also included at i$^{\prime}$ band (dashed black lines). Similar to the water clouds of Fig.~\ref{fig:CloudPhases}, the increased multiple scattering within the hazes brightens the planet compared to the clear atmosphere case, leading to larger $\pi{F_{n}}$.
However, unlike the cloudy models, which show a nearly wavelength-independent increase in $\pi{F_{n}}$, the hazes absorb more light at shorter $\lambda$ and scatter more light at longer $\lambda$, thereby displaying a wavelength-dependence for $\pi{F_{n}}$. Additionally, both the $\pi{F_{n}}$ and P$_s$ curves show wavelength-dependent features for the hazes that smooth out with increasing $\lambda$, with P$_s$ being more sensitive to the addition of hazes than $\pi{F_{n}}$. While the water clouds of Fig.~\ref{fig:CloudPhases} show a single rainbow feature at $\alpha$ $\approx$ 40$\degree$ polarized perpendicular to the scattering plane, the hazy models show multiple features that are instead polarized parallel to the scattering plane and whose locations shift in $\alpha$ with changing $\lambda$. We acknowledge, however, that these shifting features, especially the multiple undulations at shorter $\lambda$ (e.g., green line in Fig.~\ref{fig:ArchHaze}), could be features caused by the Mie-scattering spherical particles used for these model hazes (see Section~\ref{sec: CloudsHazes}) and therefore might not be observed in an exoplanet that contains non-spherical haze particles. We discuss the effects of using Mie vs. non-spherical scattering haze particles in more detail in Section~\ref{sec: MievsFrac}.

\section{Impact of Non-spherical Particles and Different Biomats} \label{sec: Realistic}

Our models of the Earth Through Time have so far assumed spherical (i.e., Mie-scattering) aerosol particles. However, for many naturally irregular particles such as those in ice clouds and hazes, Mie-scattering is a simplification. Simplifications such as these are typical for theoretical studies focused on modeling exoplanets \citep[e.g.,][]{lunamorley2021}, and can greatly decrease computational runtime and storage requirements, but can also have noticeable effects on the models \citep[e.g.,][]{feng2018}. Similarly, different surface albedos can also affect the resulting signatures of the model planets \citep[e.g.,][]{gordon2023}.
In this section we explore the influence of non-spherical cloud and haze particles as well as different microbial surfaces on the resulting $\pi{F_{n}}$ and P$_s$ of our Earth Through Time models.

\subsection{Mie vs. Non-spherical Scattering in Clouds \& Hazes} \label{sec: MievsFrac}

In Section~\ref{sec: Results} we used Mie theory (i.e., spherical particles) to calculate the scattering properties of our cloud and haze particles.
However, in reality H$_2$O ice clouds are composed of crystals of varying shapes and sizes depending on the temperature and humidity of their environment \citep[e.g.,][]{magono1966, heymsfield1984}. Additionally, hydrocarbon haze is composed of chains of smaller monomer molecules that link and clump together to form aggregates \citep[e.g.,][]{bar1988, cabane1993}. The optical properties of both of these particles can be better approximated using non-spherical scattering models \citep[e.g.,][]{mishchenko2000}. Non-spherical particles tend to produce less extinction and be more forward-scattering in the VNIR compared to equal-mass spherical particles \citep[see, e.g.,][]{arney2016, arney2017, wolf2010}.

\begin{figure}[ht!]
    \centering
    \includegraphics[width=\linewidth]{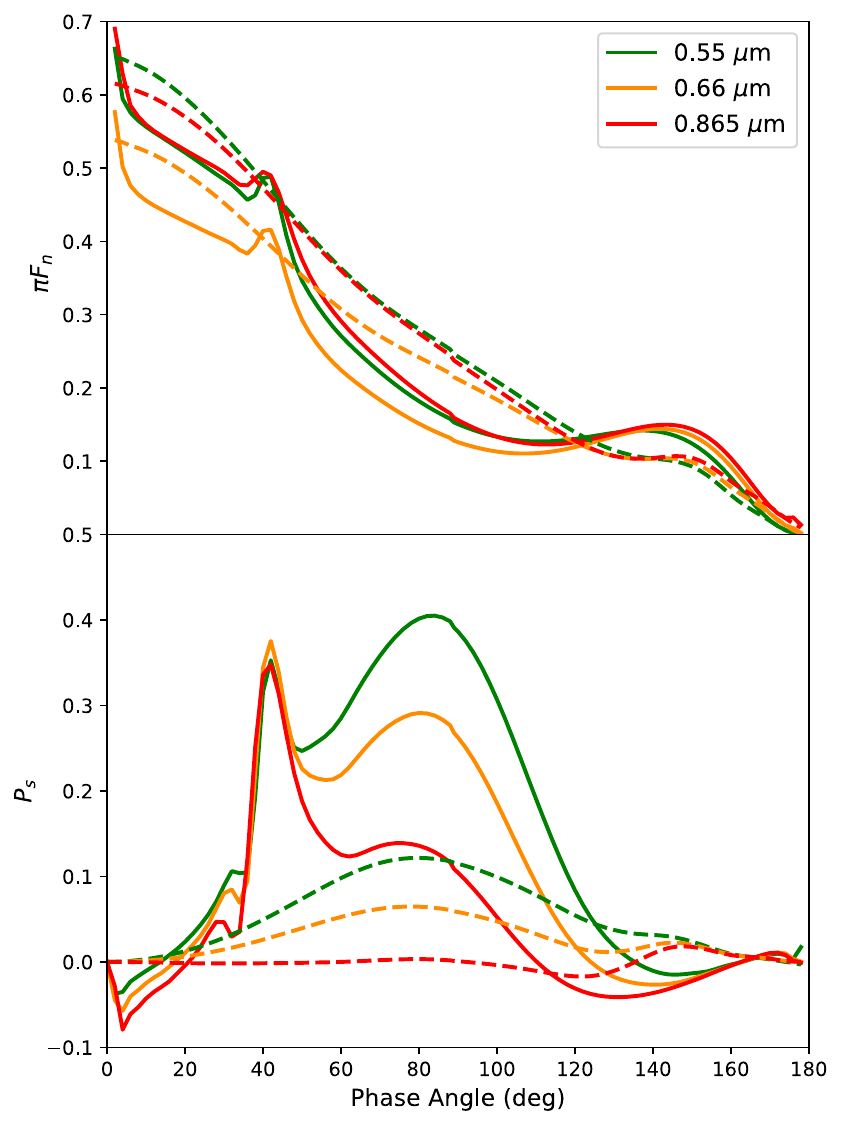}
    \caption{$\pi{F_{n}}$($\alpha$) (top) and P$_s$($\alpha$) (bottom) for a homogeneous planet with an ocean surface and Modern Earth atmosphere containing Cirrus clouds, at $\lambda$ = 0.55 $\mu$m (green lines), 0.66 $\mu$m (orange lines), and 0.865 $\mu$m (red lines). The clouds are made of either spherical (solid lines) or hexagonal (dashed lines) particles.}
    \label{fig:MievsFracIce}
\end{figure}

Fig.~\ref{fig:MievsFracIce} compares $\pi{F_{n}}$($\alpha$) and P$_s$($\alpha$) between models of Modern atmospheres with a depolarizing Lambertian ocean surface (see Section~\ref{sec: Surfs} and Fig.~\ref{fig:surfsmain}) and Cirrus clouds, whose particles were generated either with Mie scattering (solid lines) or non-spherical scattering (dashed lines). The cloud particles for the latter models were generated using scattering matrices of imperfect hexagonal ice polycrystals whose sizes range from $\sim$6 $\mu$m to $\sim$2 mm, at wavelengths of 0.55 (green lines), 0.66 (orange lines), and 0.865 $\mu$m (red lines) \citep[see][and references therein for more details]{hess1998, karalidi2012rainbow}.

Due to increased scattering off of and within the hexagonal crystals, the non-spherical ice clouds reflect more light than the Mie-scattering ice clouds, especially near $\alpha$ $\approx$ 90$\degree$ \citep[see][their Fig. 3 and Fig. 4, for the single scattering phase functions of the ice particles]{karalidi2012rainbow}. At larger $\alpha$, most of the reflected light has been scattered by the gas molecules in the atmospheric layers above the clouds, therefore decreasing the influence of the clouds. This leads to higher levels of background unpolarized light that lowers P$_s$ for both types of clouds. However, the larger size of the hexagonal ice crystals compared to the spherical ice particles leads to more absorption of the light that is not scattered at these larger $\alpha$, resulting in lower $\pi{F_{n}}$ and higher P$_s$ at $\alpha >$ 120$\degree$.

As expected, both types of cloudy models show Rayleigh peaks, caused by scattering by the gas molecules above the clouds, in their P$_s$($\alpha$) that decrease with increasing $\lambda$. 
Note that the Mie ice clouds, like the liquid water clouds in Fig.~\ref{fig:CloudPhases}, show a primary rainbow feature around $\alpha$ = 40$\degree$. The P$_s$ of the ice cloud rainbow is larger than that of the liquid water clouds due to the larger particle sizes \citep[e.g.,][]{gordon2023}. The hexagonal ice clouds, however, do not show this feature, in agreement with Earth observations \citep[see, e.g.,][and references therein]{karalidi2012rainbow}.

\begin{figure}[ht!]
    \centering
    \includegraphics[width=\linewidth]{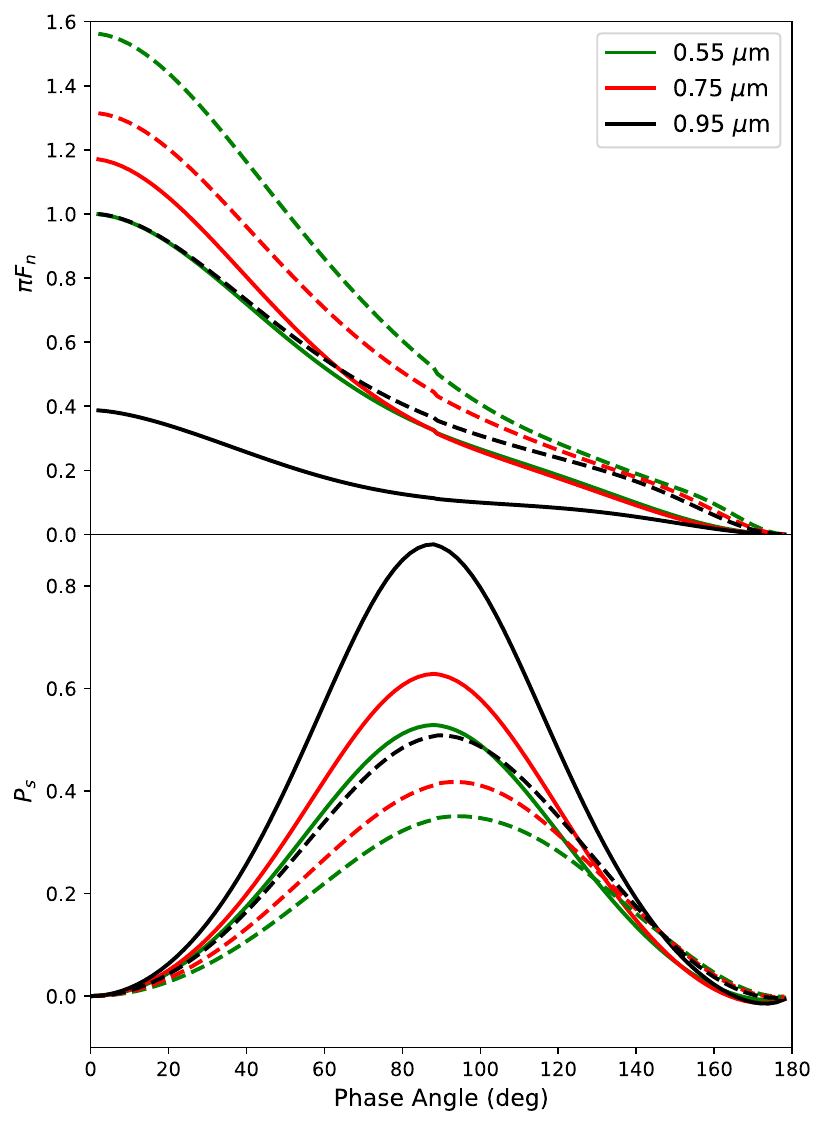}
    \caption{$\pi{F_{n}}$($\alpha$) (top) and P$_s$($\alpha$) (bottom) of a homogeneous planet with an ocean surface and Hadean: Impact atmosphere containing hydrocarbon hazes, at $\lambda$ = 0.55 $\mu$m (green lines), 0.75 $\mu$m (red lines), and 0.95 $\mu$m (black lines). The hazes are made of either spherical (solid lines) or non-spherical (dashed lines) particles.}
    \label{fig:MievsFracHaze}
\end{figure}

In Fig.~\ref{fig:MievsFracHaze} we show $\pi{F_{n}}$($\alpha$) and P$_s$($\alpha$) of an ocean planet with a Hadean: Impact atmosphere containing either Mie (solid lines) or non-spherical scattering (dashed lines) hazes.
For the non-spherical hazes, we used the haze particles of \citet[][]{karalidi2013flux}, which are modeled as randomly oriented aggregates of 94 equally sized spheres that coagulate through a cluster-cluster aggregation method into a particle with a volume-equivalent-sphere radius of 0.16 $\mu$m \citep[][]{karalidi2013flux}. The optical properties of the particles were calculated at $\lambda$ = 0.55, 0.75, and 0.95 $\mu$m using the T-matrix theory combined with the superposition theorem \citep[][]{mackowski2011}, with a wavelength-independent complex refractive index of 1.5 $\pm$ 0.001i \citep[][]{karalidi2013flux}.

Because the non-spherical particles are made of monomers, the light is scattered more within each particle in comparison to the spherical particles, therefore increasing the $\pi{F_{n}}$ but decreasing the P$_s$ from the haze compared to those of the spherical particles \citep[e.g.,][]{karalidi2013flux}. Both $\pi{F_{n}}$($\alpha$) and P$_s$($\alpha$) for both types of hazes are relatively featureless due to the small sizes of the particles, with Rayleigh scattering dominating the curves. While the disk-integrated P$_s$ of the Mie hazes peaks at $\alpha \approx$ 90$\degree$ for all three $\lambda$, the non-spherical hazes show a wavelength-dependent shift in the peak towards larger $\alpha$.

\subsection{Biomat Surface Reflectances} \label{sec: BioSurfs}

\begin{figure*}[ht!]
    \centering
    \includegraphics[width=18cm]{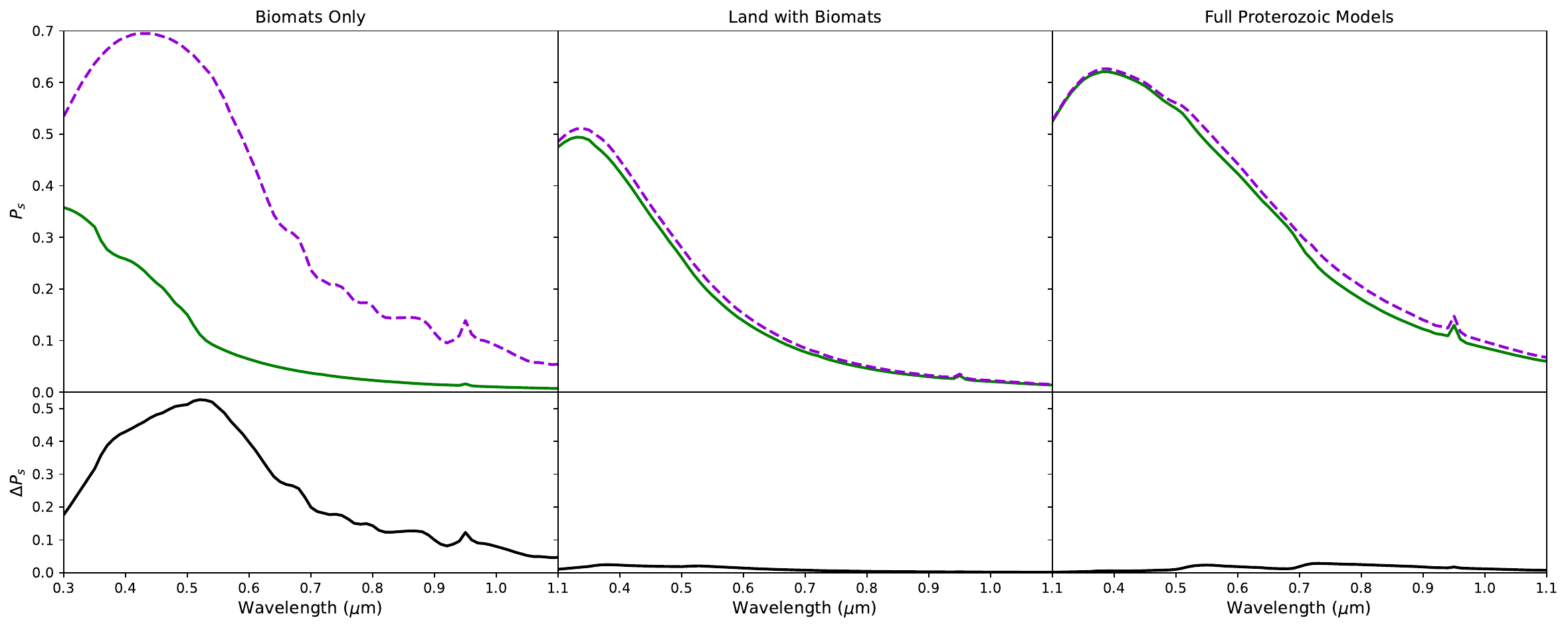}
    \caption{Comparisons at quadrature of P$_s$($\lambda$) between Proterozoic models with clear atmospheres and biological surfaces from \citet[][]{coelho2022} (solid green lines) or \citet[][]{sparks2021} (dashed purple lines). As we zoom out (from left to right) from a pixel to a Proto.~land to a full planet view, the reduction in surface microbial coverage (100\%, 5\%, and 2.1\%, respectively) results in smaller differences on the resulting disk-integrated P$_s$($\lambda$) for our weighted-averaged models. Absolute differences between the two spectra in each panel are shown in the bottom row.}
    \label{fig:BioComp}
\end{figure*}

Microbial surfaces can provide additional signatures to search for life on exoplanets \citep[e.g.,][]{schwieterman2015}. 
Here, we used two databases, those of \citet[][]{sparks2021} and of \citet[][]{coelho2022}, to test the detectability of microbes in our planetary signals.
In Fig.~\ref{fig:BioComp} we compare P$_s$($\lambda$) at $\alpha$ = 90$\degree$ for our Proterozoic models with cloud-free atmospheres and biological surfaces from either \citet[][]{coelho2022} (like those used in Section~\ref{sec: Results}; solid green lines) or \citet[][]{sparks2021} (dashed purple lines). Due to the limited wavelength range over which the biological surfaces of the latter study were measured, the models in this section only cover wavelengths from 0.3 to 1.1 $\mu$m.

At the individual pixel scale (top left panel), where the entire surface is covered by either orange and white pigments or continental microbial mats, the \citet[][]{sparks2021} model reaches P$_s$ $\approx$ 0.7, while the \citet[][]{coelho2022} model reaches P$_s$ $\approx$ 0.35. The maximum absolute difference, $\Delta$P$_s$($\lambda$), between the spectra reaches $\sim$0.53 at $\lambda$ = 0.53 $\mu$m (bottom left panel).
These differences are due to the higher albedos of the \citet[][]{coelho2022} microbial surfaces (see Fig.~\ref{fig:surfsKalt}).
At the Proto.~land scale (top middle panel; see Table~\ref{table:epochs}), the effects from the microbial surfaces on the resulting spectra decrease dramatically due to the microbes now covering only 5\% of the model surface, with a maximum $\Delta$P$_s$($\lambda$) of $\sim$0.025 at $\lambda$ = 0.38 $\mu$m (bottom middle panel).
At the full planet scale (top right panel), the maximum $\Delta$P$_s$($\lambda$) only reaches $\sim$0.028 at $\lambda$ = 0.73 $\mu$m (bottom right panel) due to the small percentage of microbes on the surface ($\sim$2.1\%) compared to, e.g., the planetary ocean (85\%). The changes in the wavelengths and values of the maximum $\Delta$P$_s$ across the three panels is due to the different combinations of surfaces in the models, with the introduction of the green pigments and marine intertidal microbial mats in the full planet models shifting the maximum $\Delta$P$_s$ to longer $\lambda$. Our results show that while the different microbial surfaces strongly influence P$_s$ of local regions on our clear atmosphere model planets, integrated across larger regions or the entire planetary disk, the microbial surfaces have a smaller influence on the resulting signal.

\section{Observing Constraints for the Next Generation Telescopes} \label{sec: Scaling}

\begin{figure*}[ht!]
    \centering
    \includegraphics[width=14.5cm]{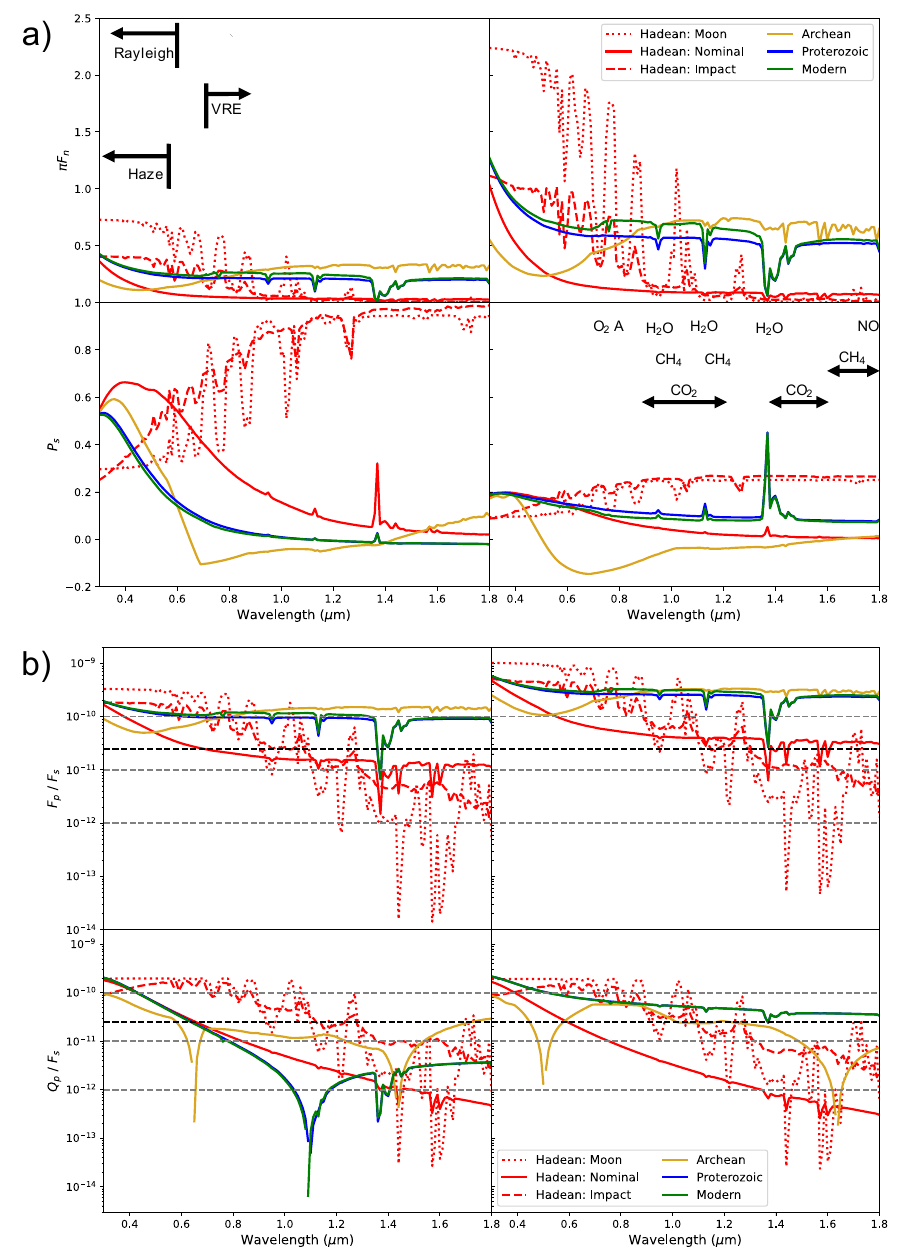}
    \caption{(a) $\pi{F_{n}}$($\lambda$) (top row) and P$_s$($\lambda$) (bottom row) for the full models of our six Earth Through Time epochs, with their atmospheres containing patchy clouds and hazes where applicable. The spectra are plotted at either quadrature (left column) or $\alpha$ = 40$\degree$ (right column). (b) Unpolarized (top row) and polarized (bottom row) contrast ratios for the six models of panel (a) scaled to their corresponding solar spectra through time. These contrast ratios assume the planets are in circular edge-on orbits. Also marked are contrast ratio limits of 1 $\times$ 10$^{-10}$, 1 $\times$ 10$^{-11}$, and 1 $\times$10 $^{-12}$ (dashed grey lines), as well as the preliminary HWO lower limit of 2.5 $\times$ 10$^{-11}$ (dashed black lines). The breaks in some of the $Q{_p} / F{_s}$ curves are due to the polarized light changing directions (see Eq.~\ref{eq:degofpol}). Note the distinguishing spectral features of different species across different epochs as well as a detectable VRE in our Modern Earth model.}
    \label{fig:scaleratios}
\end{figure*}

The National Academy of Sciences Astronomy \& Astrophysics 2020 Decadal Survey (hereafter Astro2020) recommended a new ``Great Observatories" program telescope with the priority capability of ``direct imaging to probe polarized ocean glint on terrestrial planets" \citep[][]{Astro2020}. The NASA Habitable Worlds Observatory (hereafter HWO), recently announced in response to these recommendations from Astro2020, is expected to draw heavily from the designs of proposed missions including LUVOIR \citep[e.g.,][]{luvoir2019} and HabEx \citep[e.g.,][]{gaudi2020}. However, the full performance and capabilities of HWO are yet to be determined, and plans for different instruments are still in development.

With this in mind, we discuss here the detectability of our six different Earth Through Time models around an evolving solar-type star \citep[][]{claire2012}. 
To calculate the unpolarized planet-to-star contrast ratios, $F{_p} / F{_s}$, we computed the total reflected flux, $F_p$($\lambda$, $\alpha$), from the exoplanet arriving at the observer. We then calculated the total stellar flux seen at the observer, $F_s$($\lambda$):

\begin{eqnarray}
F_s = \frac{L}{4\pi d^2}
\label{eq:fluxstar}
\end{eqnarray}
where L is the stellar luminosity calculated from the incident flux $F_0$ at our model planet at an orbital distance $a$ away from the star, such that $L = F_0 * 4\pi {a^2} $. By combining Eqs.~\ref{eq:fluxplanet} and \ref{eq:fluxstar}, we find that the unpolarized contrast ratios are given by:

\begin{eqnarray}
\frac{F_p}{F_s} = \frac{1}{4} \textbf{S}(\lambda, \alpha) \left({\frac{r}{a}}\right)^2.
\label{eq:contrastratio}
\end{eqnarray}

To obtain the polarized contrast ratios, $Q{_p} / F{_s}$, we multiply the $F{_p} / F{_s}$ for each model by the absolute value of its corresponding P$_s$ (see Eq.~\ref{eq:degofpol}). Additionally, since the majority of sun-like stars are unpolarized \citep[e.g.,][]{kemp1987, cotton2017}, observing our star-planet system in linearly polarized light (i.e., using a linear polarizer) allows us to reduce the stellar contribution by half for these polarized contrast ratios, since only half of the randomly-oriented stellar photons can pass through the linear polarizer \citep[e.g.,][]{collett2005}.
For our calculations here, we used a planetary radius $r$ = $R_{\earth}$ and an orbital distance $a$ = 1 AU. We acknowledge that these contrast ratios can also depend on the phase angle and inclination of the planet, but for simplicity, we assume circular edge-on orbits (i.e., $i$ = 90$\degree$).
Finally, following previous studies for theoretical simulations of contrast ratios of terrestrial exoplanets \citep[see, e.g.,][]{mahapatra2023, vaughan2023}, we assume here that we have a perfect removal of any starlight (including all noise) from the planetary pixel using information from neighboring pixels in the detector.

Panel (a) of Fig.~\ref{fig:scaleratios} shows $\pi{F_{n}}$($\lambda$) (top row) and P$_s$($\lambda$) (bottom row) for the full models of our six Earth Through Time epochs, while panel (b) shows the resulting unpolarized (top row) and polarized (bottom row) contrast curves for these epochs. We display here two key phase angles: $\alpha$ = 90$\degree$, to capture the planets at their widest inner working angle (IWA) (left columns), and $\alpha$ = 40$\degree$, to capture the planets at the peak of the water cloud rainbow feature (right columns). 
For the models that can possess hazes (Hadean: Impact and Archean), we assumed an equal distribution of 50\% hazy and 50\% clear atmospheres. For the models that can have water clouds (Proterozoic and Modern), we assumed that $\sim$67\% of the planet was covered by clouds \citep[e.g.,][]{king2013}, which we divided up as 34\% Cirrus clouds and 33\% Stratocumulus clouds. 
The Hadean: Moon and Hadean: Nominal models have clear atmospheres here, although in reality we would expect some cloud cover during these eras. 
Due to the limited wavelengths over which the non-spherical aerosol particles were calculated in Section~\ref{sec: MievsFrac}, our models here used the Mie-scattering aerosols of Section~\ref{sec: CloudsHazes}.
Panel (b) of Fig.~\ref{fig:scaleratios} also includes dashed grey lines denoting contrast ratio limits ranging from 1 $\times$ 10$^{-10}$ to 1 $\times$ 10$^{-12}$. Preliminary estimates of nearby target stars chosen for optimal HWO observations\footnote{Available online: \url{https://exoplanetarchive.ipac.caltech.edu/docs/2645_NASA_ExEP_Target_List_HWO_Documentation_2023.pdf}} \citep[][]{mamajek2024} suggest a lower contrast ratio limit for the telescope of 2.5 $\times$ 10$^{-11}$ (denoted by the black dashed lines in the panel (b) plots).

In unpolarized light, we are able to resolve the VRE at $\sim$0.7 $\mu$m, O$_2$ A-band at 0.76 $\mu$m, and NIR H$_2$O absorption at 0.93 $\mu$m at both $\alpha$ in $F{_p} / F{_s}$ for the Modern Earth scenario above a contrast of 1 $\times$ 10$^{-10}$, the raw limit set by preliminary studies for LUVOIR and HabEx \citep[e.g.,][]{luvoir2019, gaudi2020}.
At $\alpha$ = 90$\degree$, characterizing any additional absorption bands in the NIR requires a contrast ratio below this limit. Additionally, all three of the habitable planet scenarios (Archean, Proterozoic, and Modern) have $F{_p} / F{_s}$ above the HWO contrast limit at $\alpha$ = 90$\degree$, save for the full depth of the 1.35 $\mu$m H$_2$O feature.
The hotter, highly-pressurized, and thus more absorbing Hadean: Moon and Hadean: Impact scenarios lead to lower contrasts in the NIR (down to $\sim$1.0 $\times$ 10$^{-13}$ and $\sim$1 $\times$ 10$^{-12}$, respectively) than the other four epochs, with the deep NIR CO$_2$ and CH$_4$ absorption bands in the Hadean: Moon model requiring contrasts lower than 1 $\times$ 10$^{-13}$. However, the stronger Rayleigh scattering in these models due to CO$_2$ (for the Hadean: Moon) and H$_2$ (for the Hadean: Impact) dominating their atmospheres leads to higher contrasts in the UV and VIS (up to $\sim$3.2 $\times$ 10$^{-10}$ and $\sim$1.9 $\times$ 10$^{-10}$, respectively).
As discussed in Section~\ref{sec: clear}, all models are brighter towards smaller $\alpha$ due to a larger portion of the visible disk being illuminated. This is detectable in the resulting $F{_p} / F{_s}$ of all six models at $\alpha$ = 40$\degree$. At this phase, all three habitable planet scenarios are detectable above the LUVOIR and HabEx raw limit of 1 $\times$ 10$^{-10}$ across the full spectrum, again with the exception of the 1.35 $\mu$m H$_2$O feature. All three Hadean scenarios are now detectable above the HWO lower contrast limit in the VIS and some NIR wavelengths, with the deeper NIR CO$_2$ and CH$_4$ absorption bands still requiring lower contrasts. We acknowledge, however, that at smaller $\alpha$, the planet is closer to its host star and thus can be more difficult to observe, requiring tighter IWAs for the telescope to resolve it.

Since only a fraction of the light for each model planet becomes linearly polarized, the resulting $Q{_p} / F{_s}$ for all six epochs are lower than their corresponding unpolarized contrasts for both $\alpha$. 
As discussed in Section~\ref{sec: clear}, the Hadean: Moon and Hadean: Impact scenarios have more absorption in the NIR than the other models, thus leading to less reflected flux at longer $\lambda$. However, any reflected light that does remain is singly scattered in the upper atmosphere and becomes highly polarized, leading to more reflected flux from the planet being detected and thus increasing $Q{_p} / F{_s}$ for these two scenarios.
The breaks in $Q{_p} / F{_s}$ at $\lambda$ $\approx$ 1.1 $\mu$m for the Proterozoic and Modern scenarios at quadrature, as well as those at $\approx$ 0.65 $\mu$m and $\approx$ 1.45 $\mu$m (for $\alpha$ = 90$\degree$) and $\approx$ 0.5 $\mu$m and $\approx$ 1.65 $\mu$m (for $\alpha$ = 40$\degree$) for the Archean scenario, are not due to absorptions but rather to a change in the direction of the polarization (see Eq.~\ref{eq:degofpol}), with the light switching from being polarized perpendicular (for shorter $\lambda$s) to parallel (for larger $\lambda$s) to the scattering plane. This is due to the introduction of the aerosols in these models, which changes the polarization state of the planets. Although the Hadean: Impact scenario also includes patchy hydrocarbon hazes here, the smaller $\tau$ and differing vertical distribution of particle radii modes (see Section~\ref{sec: CloudsHazes}) for the hazes in this model compared to those in the Archean scenario do not cause a change in the direction of the polarization. While the polarized contrasts for the Proterozoic and Modern models drop below 1 $\times$ 10$^{-12}$ at $\alpha$ = 90$\degree$ due to scattering by the liquid water clouds (see Section~\ref{sec: cloudhaze}), these clouds create a peak in P$_s$ at $\alpha$ $\approx$ 40$\degree$. At this phase, the $Q{_p} / F{_s}$ of these models are now detectable above the HWO contrast limit across the full spectrum.

Our calculated contrast ratios provide important preliminary predictions for the unpolarized and polarized planet-to-star flux ratios required to characterize Earth-like exoplanets across all evolutionary stages. Our results suggest that an Earth-like planet at any point in its history, from a young and hot Hadean-like planet to a current Modern-like planet, could be characterized around a Sun-like star in unpolarized (polarized) light if the contrast ratio capability of future instruments could be pushed to a lower limit of 1 $\times$ 10$^{-12}$ (1 $\times$ 10$^{-13}$).

\section{Discussion And Conclusions} \label{sec: DiscussConclude}

Earth has gone through many geological and ecological changes throughout its history, with evidence of life existing on the surface as early as the Archean eon. Comparing observations of Earth-like exoplanets only to the modern Earth therefore limits our characterizations of these worlds and our ability to assess their habitability. Previous studies that focused on the changes in Earth's spectra through its evolution provided useful and important analyses but missed out on the full informational content available in the reflected light. 
In this work we presented the first numerical simulations of the spectropolarimetric signals of the Earth throughout its history, including, to our knowledge, the first unpolarized and polarized models of the spectra of the planet during the Hadean eon. Our models cover the full VNIR wavelength range ($\lambda$ = 0.3 - 1.8 $\mu$m, $\Delta$$\lambda$ = 10 nm) and the full phase angle space ($\alpha$ = 0$\degree$ - 180$\degree$, in 2$\degree$ steps). With the planned development of near-future polarimeters for both ground and space-based observatories, including the upcoming ELTs and HWO, as well as increased interest in polarization among the astrophysical community, we expect that these models can provide valuable feedback to the community. All of our Earth Through Time models are publicly available online through Zenodo: \dataset[doi:10.5281/zenodo.13882511]{\doi{10.5281/zenodo.13882511}}.

For the clear (i.e., cloud- and haze-free) atmosphere planets modeled in Section~\ref{sec: Results}, differences in the surface reflectivities and atmospheric VMRs across the different epochs allow us to distinguish between the habitable and non-habitable scenarios (see Fig.~\ref{fig:clearspectra}). The hotter atmospheres and higher surface pressures of the two short-lived, non-habitable Hadean scenarios cause increased absorption in the NIR, leading to lower $\pi{F_{n}}$ but higher P$_s$ compared to the habitable scenarios. Additionally, while the Hadean: Nominal model produces similar $\pi{F_{n}}$ and P$_s$ to the habitable scenarios, the lower levels of H$_2$O and higher levels of CO in its atmosphere compared to the other scenarios leads to distinguishing spectral features. The higher amounts of CH$_4$ in the Archean atmosphere and O$_2$ in the Modern atmosphere lead to detectable NIR absorption bands in the former and an O$_2$ A-band in the latter, thereby allowing for further differences between the habitable scenarios themselves.

The addition of clouds and hazes to our models flattens the resulting total flux and polarization spectra and reduces absorption and surface features, especially in P$_s$ (see bottom panels of Fig.~\ref{fig:ProtoClouds} and Fig.~\ref{fig:HadImpHaze}). The hydrocarbon hazes produce additional wide absorption features in the UV and VIS that darken the planets (i.e., lower $\pi{F_{n}}$, see top panel of Fig.~\ref{fig:HadImpHaze}) and invert the direction of the polarization (see bottom panel of Fig.~\ref{fig:HadImpHaze}). Additionally, while the spectra provide information on some absorption features for the cloudy and hazy models, phase curves provide important defining characteristics of the aerosols themselves, including the distinctive $\alpha$ $\approx$ 40$\degree$ rainbow feature of the liquid water cloud droplets (see Fig.~\ref{fig:CloudPhases}). 
These aerosol features are more prominent in P$_s$ than $\pi{F_{n}}$ for our models. Our results therefore highlight the importance of using both flux and polarization measurements across both wavelength and phase space to fully characterize the planets.

We acknowledge that our simplifications of a single cloud (haze) layer per pixel and the adoption of a single cloud (haze) size distribution across the planetary disk can affect the overall shapes of our model spectra and phase curves \citep[e.g.,][]{karalidi2012rainbow, gordon2023}. In reality, cloud particle sizes vary significantly across Earth, as do hazes across planets such as Jupiter and Titan, and these planets show overlap of clouds (hazes) with different properties \citep[see, e.g.,][and references therein]{han1994, hess1998, atreya2005, west2015}. \citet[][]{karalidi2012rainbow} modeled the Earth with overlapping layers of liquid water clouds of different size distributions, as well as liquid water clouds covered by ice clouds, and showed, for example, that the ice cloud coverage did not mask the liquid water rainbow feature at $\alpha$ $\approx$ 40$\degree$. 
For more information on the effects of overlapping clouds on the resulting $\pi{F_{n}}$ and P$_s$ of terrestrial planets, we refer the reader to this study and the references within. In a future paper we plan to include models of terrestrial planets with changing amounts of clouds and hazes overlapping in the same atmospheres.

While assuming idealistic cases for atmospheric and surface models is common practice in simulating planetary signals \citep[e.g.,][]{mclean2017, tilstra2021}, too many simplifications affect the models \citep[e.g.,][]{feng2018, lunamorley2021, gordon2023}. Here we showed that incorporating non-spherical scattering cloud and haze particles (Section~\ref{sec: MievsFrac}) as well as different microbial surfaces (Section~\ref{sec: BioSurfs}) create noticeable differences in our models. We acknowledge that changing levels of other surfaces would create similar variations in the polarized signals. The non-spherical clouds and hazes removed extraneous 
features in the resulting signals that were created by the Mie-scattering aerosols.
Incorporating physically consistent model parameters will therefore be crucial for characterizing future observations and retrieving the true atmospheric and surface properties of potentially habitable exoplanets. However, we acknowledge that increasing the complexity of fitted models will increase the computing power and time needed to run them, so understanding which simplifications can be made to models in different scenarios without loss of necessary information will be vital.

To determine when we could detect noticeable differences in the disk-integrated P$_s$ between spectra with the different microbial databases, we ran a preliminary parameter scan of models with microbial surface coverages ranging from 5\% to 95\% (in steps of 5\%). We acknowledge that different combinations of surfaces can alter the levels of P$_s$ (see Fig.~\ref{fig:BioComp}), but for simplicity the models in this scan only used microbial and ocean albedos, with all land and coast surfaces covered completely by microbes. Any increase in the microbial coverage was balanced by a corresponding decrease in the ocean coverage. The models here used our clear Proterozoic Earth atmosphere and we therefore did not investigate the effects of clouds on these signals.
For each model we then calculated the signal-to-noise ratio (SNR) required to detect the planet at quadrature around a Sun-like star, following the standard CCD equation and assuming a distance d = 10 pc from the observer. For these simplified calculations we did not take into account zodiacal or exozodiacal light, but instead assumed a background noise based on the JWST background model\footnote{\url{https://jwst-docs.stsci.edu/jwst-general-support/jwst-background-model}} at a reference wavelength of 0.64 $\mu$m \citep[][]{rigby2023}. In the absence of defined detector performance parameters for the LUVOIR or HabEx mission studies, we used the quantum efficiency, readout noise, and dark current noise of the HST Wide Field Camera 3 instrument, which is optimized for observations in the VIS range \citep[e.g.,][]{marinelli2024}.
At a total exposure time limit of 60 days, the LUVOIR and HabEx studies define their preferred SNR for spectra as 20 per resolution element, with SNR = 10 being acceptable \citep[see, e.g.,][and references therein]{mamajek2024}. We therefore define a detectable difference between our models here with the different microbial databases as an absolute difference ($\Delta$SNR) of 10 in the SNR between the spectra.

We found that a cloud-free planet covered by 20\% microbes produced a $\Delta$SNR $>$ 10 between the two spectra across the full VNIR range, while a planet covered by 25\% microbes produced a $\Delta$SNR $>$ 20 for all $\lambda$ $>$ 0.4 $\mu$m.
While our parameter scan and SNR calculations were simplified and non-exhaustive, our results suggest that if an exoplanet observed at quadrature had approximately a fifth or more of its surface covered by microorganisms, similar to the total coverage of vegetation and cropland on modern-day Earth \citep[see, e.g.,][]{friedl2002, friedl2010}, we could expect to see spectral differences between the different pigments of life. We acknowledge, however, that this surface test, as well as the models discussed in Section \ref{sec: BioSurfs}, were performed with cloud-free pixels only, and that the inclusion of clouds above the microbial surfaces would affect the resulting signals and calculated SNRs. For example, the inclusion of Cirrus clouds in the atmospheres of these pixels (not shown here, but models are available in Zenodo) reduces the detectability by $\sim$50\% across the spectra, requiring a cloudy planet to be covered by at least 30\% microbes to obtain a $\Delta$SNR $>$ 10.

We also emphasize that the disk-integrated signals of our model planets were simulated using the weighted-averaging method and thus are only based on percent mixtures of pixels with different surfaces. Therefore, while our models give a general idea of the observed signal, we miss the directionality of the reflected light, which can further alter the signals, especially for polarization \citep[see, e.g.,][]{karalidistam2012}. For example, a continental surface close to the visible disk equator that is covered in microbes or plants (e.g., the Amazon rainforest) would contribute more to the disk-integrated signal and could dominate the spectrum, even though its surface coverage is small in comparison to the surrounding ocean surface. Examining this effect on the spectra of the Earth Through Time and investigating true heterogeneity of the Earth Through Time is part of ongoing work.

Throughout this paper we demonstrated that polarimetric observations are a valuable tool complementing flux observations in the characterization of terrestrial exoplanets. Many features of our Earth Through Time models could be characterized in both unpolarized and polarized light around a Sun-like star above the preliminary HWO lower contrast limit of 2.5 $\times$ 10$^{-11}$ (see Section~\ref{sec: Scaling}). Our scattered light spectra were more distinguishable in polarization than in flux, for both clear and cloudy cases with, e.g., the VRE of our clear-atmosphere Modern Earth spectra varying only by $\sim$0.04 in $\pi{F_{n}}$ (see top panel of Fig.~\ref{fig:clearspectra}) but by $\sim$0.11 in P$_s$ (see bottom panel of Fig.~\ref{fig:clearspectra}). Additionally, polarization better differentiated our Proterozoic Earth models with different cloud coverages than flux (see Fig.~\ref{fig:CloudPhases}). Therefore, even though polarimetry requires achieving lower contrast ratios (see bottom row of panel (b) of Fig.~\ref{fig:scaleratios}) for most of the models, the diagnostic ability of polarized light highlights the importance of achieving these smaller contrasts. Achieving a contrast limit of $\sim$2 $\times$ 10$^{-11}$, just 1.25 times dimmer than the preliminary HWO lower contrast limit, would allow us to resolve all major biomarkers for our habitable planet scenarios in the visible as well as place upper limits in the NIR. This would allow us to characterize the clouds and some major atmospheric and surface components of terrestrial exoplanets. In order to resolve all absorption features across the full VNIR, which helps constrain the atmospheric structure and chemistry, our models suggest achieving a lower contrast of 1 $\times$ 10$^{-13}$.

Our models here focused on planets orbiting at 1 AU around a Sun-like star, when in reality habitable planets could exist around stars over a range of stellar types in HZs of varying semi-major axes \citep[see, e.g.,][and references therein]{hill2023, mamajek2024}. These differing orientations would alter the resulting contrast ratios of the planets and lead to different IWAs at which their features could be detected. \citet[][]{vaughan2023} provided analyses of IWAs and contrast ratios required by HWO for characterizations of specific features of modern-day Earth-like exoplanets around stars of multiple stellar types in the preliminary HWO target list, but these authors focused on observations at $\lambda$ = 0.67 $\mu$m. Our results build upon this study by including contrast ratios across a larger VNIR spectrum for multiple phase angles and for Earth-like planets across geologic time. While future direct imaging studies will be optimized for viewing planets at quadrature, both our results here and those of \citet[][]{vaughan2023} highlight the importance of looking at phase angles other than $\alpha$ = 90$\degree$ to better characterize planetary features such as clouds, hazes, and different surfaces. Additionally, our results emphasize the necessity of utilizing polarization in these characterizations to obtain more information about the planets than can be provided by unpolarized measurements alone. Pushing the HWO requirements to smaller IWAs and lower contrast limits, as well as incorporating a sensitive VNIR spectropolarimeter, would allow for the full characterizations of multiple Earth-like planets across different evolutionary stages, thus paving the way for understanding the potential habitability of terrestrial exoplanets.

\acknowledgments
We would like to thank Dr. Maxwell Millar-Blanchaer for his helpful comments and advice on the observing constraints presented in this work, as well as Dr. Jeremy Bailey for his assistance with our modeling efforts.
K.G.G., Th.K., K.B., T.K., and V.M. acknowledge the support of NASA Habitable Worlds grant No. 80NSSC20K1529. K.G.G. and Th.K. acknowledge the University of Central Florida Advanced Research Computing Center high-performance computing resources made available for conducting the research reported in this paper (\url{https://arcc.ist.ucf.edu}). 
The results reported herein benefited from collaborations and/or information exchange within NASA's Nexus for Exoplanet System Science (NExSS) research coordination network sponsored by NASA's Science Mission Directorate.
Finally, we thank the anonymous referees for helpful comments and suggestions, which resulted in improvements to our manuscript.

\bibliography{EarthThroughTime}{}
\bibliographystyle{aasjournal}



\end{document}